\documentclass[11pt]{article}
\usepackage{geometry}                
\usepackage[pdftex]{color}      
\usepackage{graphicx}
\usepackage{amssymb}
\usepackage{epstopdf}
\usepackage{setspace}
\title{The Estimation of the Effective Reproductive Number from Disease Outbreak Data}

\author{ Ariel Cintr\'{o}n-Arias$^{1,5}$ 
\footnote{corresponding author: ariel@samsi.info },
Carlos Castillo-Ch\'{a}vez$^{2}$, Lu\'{i}s M. A. Bettencourt$^3$,\\
Alun L. Lloyd $^4$, and H. T. Banks $^5$\\ 
$^1$Statistical and Applied Mathematical Sciences Institute, \\
19 T. W. Alexander Drive, P.O. Box 14006, \\ Research Triangle Park, NC 27709-4006 \\
$^2$Department of Mathematics and Statistics, \\ Arizona State University,  
P.O. Box  871804,\\  Tempe, AZ 85287 - 1804 \\
$^3$ Theoretical Division, Mathematical Modeling and Analysis (T-7),\\
Los Alamos National Laboratory, Mail Stop B284,\\ 
Los Alamos, NM 87545 \\ 
$^4$ Biomathematics Graduate Program and Department of Mathematics,\\
North Carolina State University, Raleigh, NC 27695 \\
$^5$ Center for Research in Scientific Computation, \\
North Carolina State University,
P.O. Box 8205, \\ Raleigh, NC 27695}

\date{April 22, 2008}                                           

\begin{document}
\maketitle
\pagebreak

\abstract{We consider a single outbreak susceptible-infected-recovered (SIR)
model and corresponding estimation procedures for the
effective reproductive number $\mathcal{R}(t)$. We discuss the
estimation of the underlying SIR parameters with a
generalized least squares (GLS) estimation
technique. We do this in the context of appropriate statistical
models for the measurement process. We use asymptotic statistical
theories to derive the mean and variance of the limiting
(Gaussian) sampling distribution and to perform post statistical
analysis of the inverse problems. We illustrate the ideas and
pitfalls (e.g., large condition numbers on the corresponding
Fisher information matrix) with both synthetic and influenza
incidence data sets.}

\section*{Keywords}
Effective reproductive number, basic reproduction ratio, reproduction number, 
$\mathcal{R}$, $\mathcal{R}(t)$, $\mathcal{R}_0$,
parameter estimation, ordinary least squares, generalized least squares.

\section{Introduction}\label{intro}

The transmissibility of an infection can be quantified by its 
basic reproductive number, $\mathcal{R}_0$, defined as the mean number
of secondary infections seeded by a typical infective into a completely 
susceptible (naive) host population \cite{anma,dietzestr0,het}. For many simple
epidemic processes, this parameter determines a threshold: whenever
$\mathcal{R}_0>1$, a typical infective gives rise to more than one
secondary infection, leading to an epidemic. In contrast, when 
$\mathcal{R}_0<1$, infectives typically give rise to less than one
secondary infection and the prevalence of infection cannot increase.

Due to the natural history of some infections, transmissibility is
better quantified by the effective---rather than the basic---reproductive 
number.  For instance,
exposure to influenza in previous years confers some cross-immunity
\cite{couchkasel,fgb,ncwc}; the strength of this protection depends on the
antigenic similarity between the current year's strain of influenza and
earlier ones. Consequently, the population is non-naive and so it is
more appropriate to consider the effective reproductive number, 
$\mathcal{R}(t)$, a time-dependent quantity that accounts for the
population's reduced susceptibility.

Our goal is to develop a methodology for the estimation of 
$\mathcal{R}(t)$ that also provides a measure of the uncertainty in
the estimates.  
We apply the proposed methodology in the context of annual influenza 
outbreaks, focusing on data for influenza A (H3N2) viruses, which were, 
with the exception of the influenza seasons
2000-1 and 2002-3, the dominant flu subtype in the US over the 
period from 1997 to 2005 \cite{cdcflu_act,fukuda}. 

The estimation of reproductive numbers is typically an indirect process
because some of the parameters on which these numbers depend are difficult,
if not impossible, to quantify directly. A commonly used indirect approach
involves fitting a model to some epidemiological data, providing estimates
of the required parameters.

In this study we estimate the effective reproductive number by fitting 
a deterministic epidemiological model employing either an Ordinary 
Least-Squares (OLS) or a Generalized Least-Squares (GLS) estimation
scheme to obtain estimates of model parameters. Statistical asymptotic 
theory \cite{advgil,sebwil} and sensitivity analysis \cite{cruz, saltelli} 
are then applied to give approximate sampling distributions for the
estimated parameters. Uncertainty in the estimates of $\mathcal{R}(t)$
is then quantified by drawing parameters from these sampling distributions,
simulating the corresponding deterministic model and then calculating effective
reproductive numbers. In this way, the sampling distribution of the effective 
reproductive number is constructed at any desired time point.

The statistical methodology provides a framework within which the
adequacy of the parameter estimates can be formally assessed for a given
data set. We shall
present instances in which the fitted model appears to provide an adequate
fit to a given data set but where the statistics reveal that the 
parameter estimates have very high levels of uncertainty. We also discuss
situations in which the fitted model appears, at least visually, to
provide an adequate fit and where the statistics suggest that the 
uncertainty in the parameters is not so large but that, in reality, a poor
fit has been achieved. We discuss the use of residuals plots as a diagnostic
for this outcome, which highlights the problems that arise when the
assumptions of the statistical model underlying the estimation framework
are violated.

This manuscript is organized as follows:  In Section \ref{data} the data 
sets are introduced.  A single-outbreak deterministic model is introduced 
in Section \ref{sir_mdl}.  Section \ref{est_mthds} introduces the 
least squares estimation methodology used to estimate values for the 
parameters and quantify the uncertainty in these estimates. Our methodology
for obtaining estimates of $\mathcal{R}(t)$ and its uncertainty is also
described. Use of these schemes is illustrated in Section \ref{synt_data},
in which they are applied to synthetic data sets. Section \ref{rslts_ols}
applies the estimation machinery to the influenza incidence data sets. 
We conclude with a discussion of the methodologies and their application
to the data sets.


\section{Longitudinal Incidence Data}\label{data}

\begin{table}
\caption{Number of tested specimens and influenza isolates during several annual outbreaks
in the US \cite{cdcflu_act}.}
\begin{center}
\begin{tabular}{|c|c|c|c|c|}\hline
Season & Total number of & Number of  A(H1N1) \&& Number of & Number of \\
	& tested specimens & A(H1N2) isolates& A(H3N2) isolates& B isolates \\ \hline
1997-1998	&	99,072 &	6 &	3,241 &	102 \\ \hline
1998-1999	&	102,105 &	30 &	2,607 &	3,370 \\ \hline
1999-2000	&	92,403 &	132 &	3,640 &	77 \\ \hline
2000-2001	&	88,598 &		2,061 &	66 &	4,625 \\ \hline
2001-2002	&	100,815 &	87 &	4,420 &	1,965 \\ \hline
2002-2003	&	97,649 &	2,228&	942 &	4,768 \\ \hline
2003-2004	&	130,577 &	2 &	7,189 &	249 \\ \hline
2004-2005	&	157,759 &	18 &	5,801 &	5,799 \\ \hline
Mean	&		108,622 &	571 &	 3,488 &   2,619 \\ \hline
\end{tabular}
\end{center}
\label{isoltab}
\end{table}

\begin{figure} 		
\begin{center}
\includegraphics[width=4in]{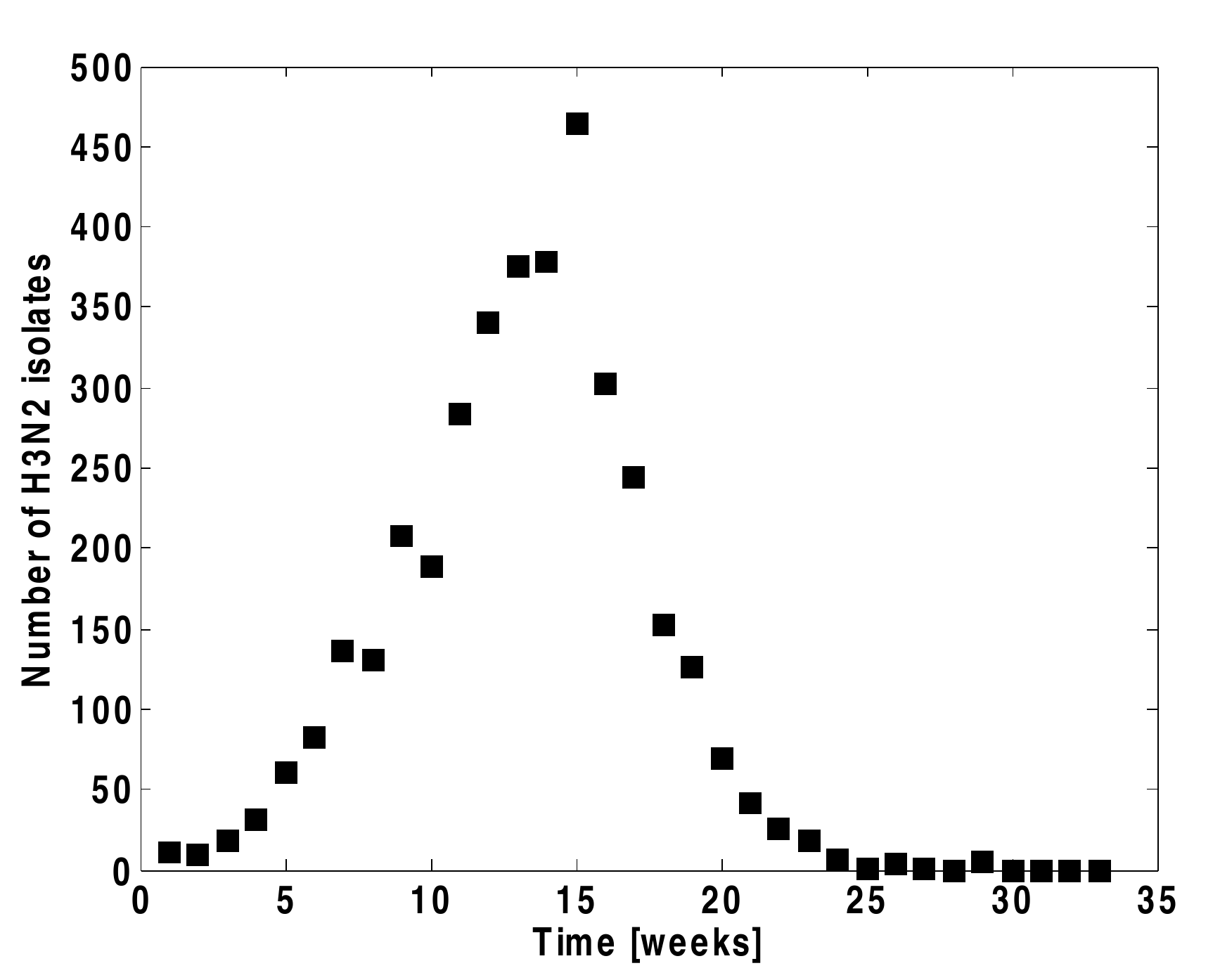}
\end{center}
\caption{Influenza isolates reported by the CDC in the US during the 1999-2000 
season \cite{cdcflu_act}.  The number of H3N2 cases (isolates) is displayed 
as a function of time.  Time is measured as the number of weeks since
the start of the year's flu season. For the 1999-2000 flu season, week 
number one corresponds
to the fortieth week of the year, falling in October.}
\label{99agst}
\end{figure}

Influenza is one of the most significant infectious diseases of humans,
as witnessed by the 1918 ``Spanish Flu'' pandemic, during which 
20 to 40 percent of the worldwide population became infected. At least
50 million deaths resulted, with 675,000 of these occurring in the 
US \cite{pandflugov}. The impact of flu is still significant during 
inter-pandemic periods: the Centers for Disease Control and Prevention
(CDC) estimate that between 5 and 20 percent
of the US population becomes infected annually \cite{cdcflu_act}. These
annual flu outbreaks lead to an average of 200,000 hospitalizations
(mostly involving young children and the elderly) and mortality that
ranges between about 900 and 13,000 deaths per year \cite{fukuda}.

The Influenza Division of the CDC reports weekly information on influenza 
activity in the US from calendar week 40 in October through week 20 in 
May \cite{cdcflu_act}, the  period referred to as the influenza season.  
Because the influenza virus exhibits a high degree of genetic variability,
data is not only collected on the number of cases but also on the 
types of influenza viruses that are circulating. A sample of viruses
isolated from patients undergoes antigenic characterization, with the
type, subtype and, in some instances, the strain of the virus being 
reported \cite{cdcflu_act}.  

The CDC acknowledges that, while these reports may help in mapping influenza 
activity (whether or not it is increasing or decreasing) throughout the US, 
they do not provide enough information to calculate how many people became 
ill with influenza during a given season. The CDC's caution likely reflects 
the uncertainty associated with the sampling process that gives rise to the
tested isolates, in particular that this process is not sufficiently 
standardized across space and time. We return to discuss this point later
in this paper.

Despite the cautionary remarks by the CDC we use these isolate 
reports as illustrative data sets to which we can apply our proposed
estimation methodologies. Interpretation of the results, however, should
be mindful of the issues associated with the data.  The total
number of tested specimens and isolates through various seasons are 
summarized in Table \ref{isoltab}.  It is observed that H3N2 viruses 
predominated in most seasons with the exception of 2000-1 and 2002-3.  
Consequently, we focus our attention on the H3N2 subtype. Figure
\ref{99agst} depicts the number of H3N2 isolates reported over the
1999-2000 influenza season.

\section{Deterministic Single-Outbreak SIR Model}\label{sir_mdl}

The model that we use is the standard Susceptible-Infective-Recovered (SIR)
model (see, for example, \cite{anma,brca}). The state variables $S(t)$, $I(t)$,
and $R(t)$ denote the number of people who are susceptible, infective, and 
recovered, respectively, at time $t$. 
It is assumed that newly infected individuals immediately
become infectious and that recovered individuals acquire permanent
immunity. The influenza season, lasting nearly 32 weeks \cite{cdcflu_act}, 
is short compared to the average lifespan, so we ignore demographic
processes (births and deaths) as well as disease-induced fatalities
and assume that the total population size remains constant. The model is    
given by the following set of nonlinear differential equations
\begin{eqnarray}
\label{sirprochp6}
\frac{dS}{dt}&=&-\beta S\frac{I}{N}\\
\label{iequation}
\frac{dI}{dt}&=&\beta S\frac{I}{N}-\gamma I\\
\frac{dR}{dt}&=&\gamma I.
\end{eqnarray}
Here, $\beta$ is the transmission parameter and $\gamma$ is the 
(per-capita) rate of recovery, the reciprocal of which gives the
average duration of infection. Notice that one of the differential equations 
is redundant because the three compartments sum to the constant population 
size: $S(t)+I(t)+R(t)=N$. We choose to track $S(t)$ and $I(t)$. The initial 
conditions of these state variables are denoted by $S(t_0)=S_0$ and 
$I(t_0)=I_0$.

The equation for the infective population (\ref{iequation}) can be rewritten as
\begin{eqnarray}
\label{infequation}
\frac{dI}{dt}= \gamma ( \mathcal{R}(t) - 1 ) I,
\end{eqnarray}
where $\mathcal{R}(t)=\frac{S(t)}{N} \mathcal{R}_0$ and 
$\mathcal{R}_0=\beta/\gamma$.
 $\mathcal{R}(t)$ is known as the effective reproductive number, while
 $\mathcal{R}_0$ is known as the basic reproductive number.  We have
that $\mathcal{R}(t)\leq\mathcal{R}_0$, with the upper bound---the
basic reproductive number---only being achieved when the entire 
population is susceptible.

We note that $\mathcal{R}(t)$ is the product of the per-infective rate at which
new infections arise and the average duration of infection, and so
the effective reproductive number gives the average number of secondary
infections caused by a single infective, at a given susceptible fraction.
The prevalence of infection increases or decreases according to whether 
$\mathcal{R}(t)$ is greater than or less than one, respectively.
Because there is no replenishment of the susceptible pool in this SIR model,
$\mathcal{R}(t)$ decreases over the course of an outbreak as susceptible 
individuals become infected.

\section{Estimation Schemes}\label{est_mthds}

In order to calculate $\mathcal{R}(t)$, it is necessary to know the two
epidemiological parameters $\beta$ and $\gamma$, as well as the
number of susceptibles, $S(t)$, and the population size, $N$. As mentioned
before, difficulties in the direct estimation of $\beta$---whose value
reflects the rate at which contacts occur in the population and the
probability of transmission occurring when a susceptible and infective
meet---and direct estimation of $S(t)$ preclude direct estimation 
of $\mathcal{R}(t)$. As
a result, we adopt an indirect approach, which proceeds by first finding 
the parameter set for which the model has the best agreement with the data
and then calculating $\mathcal{R}(t)$ by using these parameters and the
model-predicted time course of $S(t)$. Simulation of the model also requires 
knowledge of the initial values, $S_0$ and $I_0$, which must also be
estimated.

Although the model is framed
in terms of the prevalence of infection $I(t)$, the time-series data 
provides information on the weekly incidence of infection, which, in terms of
the model, is given by the integral of the rate at which new infections
arise over the week: $\int \beta S(t) I(t)/N\,dt$. We notice that the
parameters $\beta$ and $N$ only appear (both in the model and in the
expression for incidence) as the ratio $\beta/N$, precluding their separate
estimation. Consequently we need only estimate the value of this ratio, which
we call  $\tilde\beta=\beta/N$.

We employ inverse problem methodology to obtain estimates of the vector 
$\theta=(S_0,I_0,\tilde\beta,\gamma)\in \mathbb{R}^p=\mathbb{R}^4$ 
by minimizing the difference between
the model predictions and the observed data, according to two related but
distinct least squares criteria, ordinary least squares (OLS) and generalized
least squares (GLS). In what follows, we refer to $\theta$ as the parameter 
vector, or simply the parameter, in the inverse problem, even though some 
of its components are initial conditions, rather than parameters, of the 
underlying dynamic model.

\subsection{Ordinary Least Squares (OLS) Estimation}\label{ols_est}

The least squares estimation methodology is based on the {\it mathematical 
model\/} as well as a {\it statistical model\/}
for the observation process (referred to as the case counting
process). It is assumed that our known model, together with a particular
choice of parameters--- the ``true'' parameter vector, written as 
$\theta_0$---exactly describes the epidemic process, but that the 
$n$ observations, $Y_j$, are affected by random deviations (e.g., measurement
errors) from this underlying process. More precisely, it is assumed that 
\begin{equation}\label{statmodlols}
	\begin{array}{lccr}
	Y_{j}=z(t_j;\theta_0)+\epsilon_j&&& \mbox{  for $j=1,\dots,n$}
	\end{array}
\end{equation}	
where $z(t_j;\theta_0)$	denotes the weekly incidence given by the model
under the true parameter, $\theta_0$, and is defined by the following 
integral:
\begin{equation}\label{inctotal}
	z(t_j;\theta_0)=\int_{t_{j-1}}^{t_{j}}\tilde\beta S(t;\theta_0)I(t;\theta_0)\,dt.
\end{equation}
Here, $t_0$ denotes the time at which the epidemic observation process
started and the weekly observation time points are written as $t_1<\dots<t_n$. 

The errors, $\epsilon_j$, are assumed to be independent and identically 
distributed ({\it i.i.d.\/}) random variables with 
zero mean ($E[\epsilon_j]=0$), representing measurement error as well 
as other phenomena that cause the observations to deviate from the model 
predictions $z(t_j;\theta_0)$. The {\it i.i.d.\/} assumption means that the
errors are uncorrelated across time and that each has identical variance,
which we write as  $var(\epsilon_j)=\sigma_0^2$. It is assumed that
$\sigma_0^2$ is finite. We make no further assumptions about the distribution
of the errors: in particular, we {\it do not\/} assume that they are normally
distributed. It is immediately clear that we have $E[Y_j]=z(t_j; \theta_0)$
and $var(Y_j)=\sigma_0^2$: in particular, this variance is longitudinally 
constant (i.e, across the time points).

For a given set of observations $Y=(Y_1,\dots,Y_n)$, we define the 
estimator $\theta_{OLS}$ as follows:
\begin{equation}\label{olsdef}
	\theta_{OLS}(Y)=\theta_{OLS}^n(Y)=
	\mbox{arg}\min_{\theta\in\Theta}\sum_{j=1}^{n}\left[Y_j-z(t_j;\theta)\right]^2.
\end{equation}
Here $\Theta$ represents the feasible region for the parameter values. (We
discuss this region in more detail later.)
This estimator is a random variable (note that $\epsilon_j=Y_j-z(t_j;\theta_0)$
is a random variable) that involves minimizing the distance between the data 
and the model prediction. We note that all of  the observations are treated 
as having equal importance in the OLS formulation.  

If $\{y_j\}_{j=1}^n$ is a realization of the case counting (random) process 
$\{Y_j\}_{j=1}^n$, we define the cost function by
\begin{equation}
J(\theta)=\sum_{j=1}^n\left[y_j-z(t_j;\theta)\right]^2
\end{equation}
and observe that the solution of
\begin{equation} \label{thetahatols}
	\hat \theta_{OLS}=\hat\theta_{OLS}^n=\mbox{arg}\min_{\theta\in\Theta}J(\theta)
\end{equation}
provides a realization of the random variable $\theta_{OLS}$.  

The optimization problem in Equation (\ref{thetahatols}) can, in principle, 
be solved by a wide variety of algorithms. The results discussed in this
paper were obtained by using a direct search method, the Nelder-Mead simplex 
algorithm, as discussed by \cite{lagreewri}, employing the implementation
provided by the MATLAB (The Mathworks, Inc.) routine \texttt{fminsearch}.

Because $var(\epsilon_j)=E(\epsilon^2_j)=\sigma_0^2$, the true variance 
satisfies
\begin{equation}\label{truevar}
	\sigma_0^2=\frac{1}{n}E\left[\sum_{j=1}^n\left[Y_j-z(t_j;\theta_0)\right]^2\right].
\end{equation}
Because we do not know $\theta_0$, we base our estimate of the error variance 
on an equation of this form, but instead of using $\theta_0$ we
use the estimated parameter vector, $\hat\theta_{OLS}$. The right
side of Equation (\ref{truevar}) is then equal to $J(\hat\theta_{OLS})/n$. 
This estimate, however, is biased and so instead the following bias-adjusted 
estimate is used
\begin{equation}
	\hat\sigma^2_{OLS}=\frac{1}{n-4}J(\hat\theta_{OLS}).
\end{equation}
Here the $n-4$ arises because $p=4$ parameters have been
estimated from the data.

Even though the distribution of the errors is not specified, asymptotic 
theory can be used to describe the distribution of the random variable 
$\theta_{OLS}$ \cite{banksdav,sebwil}. Provided that a number of regularity
conditions as well as sampling conditions are 
met (see \cite{sebwil} for details), it can be shown that,
asymptotically (i.e., as $n\rightarrow\infty$), $\theta_{OLS}$ is distributed 
according to the following multivariate normal distribution:
\begin{equation}
\theta_{OLS}=\theta^n_{OLS}\sim\mathcal{N}_4\left(\theta_0,\Sigma_0^n\right),
\end{equation}
where
$\Sigma_0^n=n^{-1}\sigma^2_0\Omega_0^{-1}$ and
\begin{equation}
	\Omega_0=\lim_{n\rightarrow\infty}\frac{1}{n}\chi(\theta_0,n)^T\chi(\theta_0,n).
\end{equation}
We remark that the theory requires that this limit exists and that the
matrix $\Omega_0$ be non-singular. The matrix $\Sigma_0^n$ is the $4\times 4$ 
covariance matrix, whose entries equal 
$cov\left((\theta_{OLS})_i,(\theta_{OLS})_j\right)$,
and the $n\times 4$ matrix $\chi(\theta_0,n)$ is the sensitivity matrix
of the system, as defined and discussed below.

In general, $\theta_0$, $\sigma_0^2$, and $\Sigma_0^n$ are unknown, so these
quantities are approximated by the estimates $\hat\theta_{OLS}$ and 
$\hat\sigma^2_{OLS}$, and the following matrix
\begin{equation}
	\Sigma_0^n \approx \hat\Sigma_{OLS}^n = \hat\sigma_{OLS}^2\left[\chi(\hat\theta_{OLS},n)^T\chi(\hat\theta_{OLS},n)\right]^{-1}.
\end{equation}
Consequently, for large $n$, we have approximately that
\begin{equation}
\theta_{OLS}=\theta^n_{OLS}\sim\mathcal{N}_4\left(\hat\theta_{OLS},
\hat\sigma_{OLS}^2\left[\chi(\hat\theta_{OLS},n)^T\chi(\hat\theta_{OLS},n)
\right]^{-1} \right).
\end{equation}
We obtain the standard error for the $i$-th element of 
$\hat\theta_{OLS}$ by calculating 
$\sqrt{\left(\hat\Sigma^n_{OLS}\right)_{ii}}$.

The $n\times 4$ matrices $\chi(\theta,n)$ that appear in the
above formulae are called {\it sensitivity matrices\/} and are 
defined by
\begin{equation}
	\label{chi_defn}
	\chi_{ji}(\theta,n)=\frac{\partial z(t_j;\theta)}{\partial \theta_i},{\rm \qquad}  1 \le j \le n, \ \ 1\le i \le 4.
\end{equation}
The sensitivity matrix denotes the variation of the model output 
with respect to the parameter, and, for this model-based dynamical system, 
can be obtained using standard 
theory \cite{baibanks, cruz, eslami, frank, kleiber, saltelli}. 

The entries of the $j$-th row of $\chi(\theta,n)$ denote how the
weekly incidence at time $t_j$ changes in response to changes in the 
parameter (i.e., in either $S_0$, $I_0$, $\tilde\beta$, or $\gamma$) and these
can be calculated by
	\begin{eqnarray}
			\frac{\partial z}{\partial S_0}(t_j;\theta)&=&\tilde\beta\int_{t_{j-1}}^{t_{j}}\left[
			I(t;\theta)\frac{\partial S}{\partial S_0}(t;\theta)+S(t;\theta)\frac{\partial I}{\partial S_0}(t;\theta)\right]dt\\
						\frac{\partial z}{\partial I_0}(t_j;\theta)&=&\tilde\beta\int_{t_{j-1}}^{t_{j}}\left[
			I(t;\theta)\frac{\partial S}{\partial I_0}(t;\theta)+S(t;\theta)\frac{\partial I}{\partial I_0}(t;\theta)\right]dt\\
			\frac{\partial z}{\partial \tilde\beta}(t_j;\theta)&=&\int_{t_{j-1}}^{t_{j}}\left[
			S(t;\theta)I(t;\theta)+
			\tilde\beta\left(I(t;\theta)\frac{\partial S}{\partial\tilde\beta}(t;\theta)+S(t;\theta)\frac{\partial I}{\partial\tilde\beta}(t;\theta)\right)\right]dt\\
			\label{dzdgamsen}
			\frac{\partial z}{\partial \gamma}(t_j;\theta)&=&\tilde\beta\int_{t_{j-1}}^{t_{j}}\left[
			I(t;\theta)\frac{\partial S}{\partial\gamma}(t;\theta)+S(t;\theta)\frac{\partial I}{\partial\gamma}(t;\theta)\right]dt.
	\end{eqnarray}
We see that these expressions involve the partial derivatives
of the state variables, $S(t;\theta)$ and $I(t;\theta)$, with respect to the 
parameters. Analytic forms of the sensitivities are not available because 
the state variables are the solutions of a nonlinear system; instead, they 
are calculated numerically.

In order to outline how these numerical sensitivities may be found, we 
introduce the notation $x(t;\theta)=(S(t;\theta),I(t;\theta))$ and 
denote by $g=(g_1,g_2)$ the vector function whose entries
are given by the expressions on the right sides of Equations 
(\ref{sirprochp6}) and (\ref{iequation}).  Then
we can write the single-outbreak SIR model in the general vector form	
	\begin{eqnarray}\label{xvec}
		\frac{dx}{dt}(t;\theta)&=&g(x(t;\theta);\theta)\\
		x(0;\theta)&=&(\theta_1,\theta_2).
	\end{eqnarray}
Because the function $g$ is differentiable (in both $t$ and $\theta$), 
taking the partial derivatives $\partial/\partial\theta$ of both sides of 
Equation (\ref{xvec}) we obtain the differential equation
	\begin{equation}\label{dgdtheta}
		\frac{d}{dt}\frac{\partial x}{\partial \theta}=\frac{\partial g}{\partial x}\frac{\partial x}{\partial \theta}+\frac{\partial g}{\partial \theta}.
	\end{equation}
Here ${\partial g}/{\partial x} $ is a 2-by-2 matrix, 
${\partial g}/{\partial \theta}$ is a 2-by-4 matrix, and
${\partial x}/{\partial \theta}$ is the 2-by-4 matrix 
	\begin{equation}\label{dxdtheta}	
	\frac{\partial x}{\partial \theta}=
	\left[	
	\begin{array}{cccc}
		\frac{\partial S}{\partial S_0} &\frac{\partial S}{\partial I_0}&\frac{\partial S}{\partial \tilde\beta}&\frac{\partial S}{\partial \gamma}\\
		\frac{\partial I}{\partial S_0} &\frac{\partial I}{\partial I_0}&\frac{\partial I}{\partial \tilde\beta}&\frac{\partial I}{\partial \gamma}
	\end{array}
	\right].
	\end{equation}

Numerical values of the sensitivities are calculated by solving 
(\ref{xvec}) and (\ref{dgdtheta}) simultaneously.  We define 
$\phi(t)=\frac{\partial x}{\partial \theta}(t;\theta)$, let  
the parameter be evaluated at the estimate, $\theta=\hat\theta$, and
solve the following differential equations from $t=0$ to $t=t_n$

	\begin{eqnarray}
			\frac{d}{dt}x(t)&=& g(x(t;\hat\theta);\hat\theta)\\
			\frac{d}{dt}\phi(t)&=&\frac{\partial g}{\partial x}\phi(t)+\frac{\partial g}{\partial \theta}\\
			x(0)&=& (\hat\theta_1,\hat\theta_2)\\
			\phi(0)&=& \left[
				\begin{array}{cccc}
					1&0&0&0\\
					0&1&0&0		
				\end{array}\right].
	\end{eqnarray}

\subsection{Generalized Least Squares (GLS) Estimation}

	The errors in the statistical model defined by Equation 
(\ref{statmodlols}) were assumed to have constant variance, which may 
not be an appropriate 
assumption for all data sets. One alternative statistical model that can
account for more complex error structure in the case counting process
is the following
	\begin{equation}\label{stmodgls}
		Y_j=z(t_j;\theta_0)\left(1+\epsilon_j\right).
	\end{equation}
As before, it is assumed the $\epsilon_j$ are {\it i.i.d.\/} random 
variables with $E(\epsilon_j)=0$ and $var(\epsilon_j)=\sigma_0^2<\infty$, 
but no further assumptions are made. Under these assumptions, the observation 
mean is again equal to the model prediction, 
	$E[Y_j]=z(t_j;\theta_0)$, while the variance in the observations is a 
function of the time point, with $var(Y_j)=\sigma_0^2z^2(t_j;\theta_0)$. In
particular, this variance is non-constant and model-dependent. 
One situation in which this
error structure may be appropriate is when observation errors scale with
the size of the measurement (so-called {\it relative noise\/}).
	
Given a set of observations $Y=(Y_1,\dots,Y_n)$, the estimator 
$\theta_{GLS}=\theta_{GLS}(Y)$ is defined as the solution of the normal 
equations
	\begin{equation}\label{glsdef}
\sum^n_{j=1}w_j\left[Y_j-z(t_j;\theta)\right]\nabla_\theta z(t_j;\theta)=0,
	\end{equation}	
where the $w_j$ are a set of non-negative weights \cite{advgil}. 
Unlike the ordinary least squares 
formulation, this definition assigns different levels of influence, described
by the weights, to the different observations in the cost function. 
For
the error structure described above in Equation (\ref{stmodgls}), the weights
are taken to be inversely proportional to the square of the predicted
incidence: $w_j=1/[z(t_j;\theta)]^{2}$.  We shall also
consider weights taken to be proportional to the reciprocal of the 
predicted incidence; these correspond to assuming that the variance in the
observations is proportional to the value of the model (as opposed to its
square).

Suppose $\{y_j\}_{j=1}^n$ is a realization of the case counting process 
$\{Y_j\}_{j=1}^n$ and define the function $L(\theta)$ as
	\begin{equation}\label{costgls}
		L(\theta)=\sum_{j=1}^n w_j\left[y_j-z(t_j;\theta)\right]^2
	\end{equation}		
The quantity $\theta_{GLS}$ is a random variable and a realization of it, denoted by $\hat\theta_{GLS}$, is obtained by solving	
	\begin{equation}
\sum^n_{j=1}w_j\left[y_j-z(t_j;\theta)\right]\nabla_\theta z(t_j;\theta)=0,
	\end{equation}
In the limit as $n\rightarrow\infty$, the GLS estimator 
$\theta_{GLS}$ has the following asymptotic properties \cite{banksdav,advgil}:
	\begin{equation}
		\theta_{GLS}=\theta^n_{GLS}\sim \mathcal{N}_4(\theta_0,\Sigma_0^n)
	\end{equation}	
	where
	\begin{equation}
		\Sigma^n_0\approx\sigma_0^2\left[\chi(\theta_0,n)^TW(\theta_0)\chi(\theta_0,n)\right]^{-1}.
	\end{equation}
	Here $W(\theta_0)=diag(w_1(\theta_0),\dots,w_n(\theta_0))$ 
with $w_j(\theta_0)=1/[z(t_j;\theta_0)]^{2}$.  The sensitivity matrix 
$\chi(\theta_0,n)$ is as defined in Section \ref{ols_est}.
	
	Because $\theta_0$ and $\sigma_0^2$ are unknown, the estimate $\hat\theta_{GLS}$ is used to calculate approximations
	 of $\sigma_0^2$ and the covariance matrix $\Sigma_0^n$ by	 
	\begin{equation}
		\sigma_0^2\approx \hat\sigma_{GLS}^2=\frac{1}{n-4}L(\hat\theta_{GLS})
\label{sigma0_GLS}
	\end{equation}		
	\begin{equation}
		\Sigma_0^n\approx \hat\Sigma_{GLS}^n=
		\hat\sigma^2_{GLS}\left[\chi(\hat\theta_{GLS},n)^TW(\hat\theta_{GLS})\chi(\hat\theta_{GLS},n)\right]^{-1}.	
	\end{equation}	
As before, the standard errors for $\hat\theta_{GLS}$ can be approximated 
by taking the square roots of the diagonal elements of 
	the covariance matrix $\hat\Sigma^n_{GLS}$.

The cost function used in GLS estimation involves weights whose 
values depend on the values of the fitted model. These values are not
known before carrying out the estimation procedure and consequently GLS
estimation is implemented as an iterative process. An OLS is first
performed on the data, and the resulting model values provide an initial set
of weights. A weighted least squares fit is then performed using these weights,
obtaining updated model values and hence an updated set of weights. The 
weighted least squares process is repeated until some convergence criterion
is satisfied, such as successive values of the estimates being deemed to be
sufficiently close to each other. The process can be summarized as follows
	\begin{enumerate}
		\item Estimate $\hat\theta_{GLS}$ by $\hat\theta^{(0)}$ using the OLS Equation (\ref{thetahatols}).  Set $k=0$;    
		\item Form the weights $\hat w_j=1/[z(t_j;\hat\theta^{(k)})]^{2}$;
		\item Define $L(\theta)=\sum_{j=1}^n\hat w_j[y_j-z(t_j;\theta)]^2$.  Re-estimate $\hat\theta_{GLS}$ by solving
		\[
			\hat\theta^{(k+1)}=\mbox{arg}\min_{\theta\in\Theta}L(\theta)
		\]
		to obtain the $k+1$ estimate $\hat\theta^{(k+1)}$ for $\hat\theta_{GLS}$;
		\item Set $k=k+1$ and return to 2.  Terminate the procedure 
when successive estimates for $\hat\theta_{GLS}$ are sufficiently close to
each other.
	\end{enumerate}
The convergence of this procedure is discussed in \cite{carrwu,advgil}.

\subsection{Estimation of the Effective Reproductive Number}\label{rtconstr}

Let the pair $(\hat\theta,\hat\Sigma)$ denote the parameter estimate and 
covariance matrix obtained with either the OLS or GLS methodology
from a given realization $\{y_j\}_{j=1}^n$ of the case counting process.
Simulation of the SIR model then allows the time course of the susceptible
population, $S(t;\hat\theta)$, to be generated. The time course of the 
effective reproductive number can then be calculated
as $\mathcal{R}(t;\hat\theta)=S(t;\hat\theta)\hat{\tilde\beta}/\hat\gamma$.
This trajectory is our central estimate of $\mathcal{R}(t)$.

The uncertainty in the resulting estimate of $\mathcal{R}(t)$ can be assessed 
by repeated sampling of parameter vectors from the corresponding sampling 
distribution obtained from the asymptotic theory, and applying the above
methodology to calculate the $\mathcal{R}(t)$ trajectory that results each
time.
To generate $m$ such sample trajectories, we sample $m$ parameter
vectors, $\theta^{(k)}$, from the  4-multivariate normal distribution
$\mathcal{N}_4(\hat\theta,\hat\Sigma)$. We require that each $\theta^{(k)}$
lie within our feasible region, $\Theta$. If this is not the case, then we
resample until $\theta^{(k)}\in\Theta$. Numerical solution of the
SIR model using $\theta^{(k)}$ allows the sample trajectory 
$\mathcal{R}(t;\theta^{(k)})$ to be calculated. 
Below, we summarize these steps involved in the construction of the 
sampling distribution of the effective reproductive number:
	\begin{enumerate}
		\item Set $k=1$;
		\item\label{samstp} Obtain the $k$-th parameter sample from the 4-multivariate normal distribution:
			\[
				\theta^{(k)}\sim \mathcal{N}_4(\hat\theta,\hat\Sigma);
			\]
		\item If $\theta^{(k)}\notin \Theta$ (constraints are not satisfied) return to \ref{samstp}.  Otherwise go to \ref{rtcons};
		\item\label{rtcons}  Using $\theta=\theta^{(k)}$ find numerical solutions, denoted by 
		$\left(S(t;\theta^{(k)}),I(t;\theta^{(k)})\right)$, to the nonlinear system defined by
		Equations (\ref{sirprochp6}) and (\ref{iequation}).  Construct the effective reproductive number as follows:
		  	\[
				\mathcal{R}(t;\theta^{(k)})=S(t;\theta^{(k)}) \frac{\tilde\beta^{(k)} }{\gamma^{(k)}}
			\]
		where $\theta^{(k)}=\left(S_0^{(k)},I_0^{(k)},\tilde\beta^{(k)},\gamma^{(k)}\right)$;	
		\item Set $k=k+1$.  If $k>m$ then terminate, otherwise return 
to \ref{samstp}.  
	\end{enumerate}

Uncertainty estimates for $\mathcal{R}(t)$ are calculated by finding 
appropriate percentiles of the distribution of the $\mathcal{R}(t)$ samples.

\section{Estimation Schemes Applied to Synthetic Data}\label{synt_data}

\subsection{Synthetic Data with Constant Variance Noise}

We illustrate the OLS methodology and investigate its performance
 using synthetic data.  A true parameter 
$\theta_0$ is chosen and a set of synthetic data is constructed by 
adding random noise to the model prediction of incidence (for every time 
point $t_j$) in the following manner:
\begin{equation}\label{noise_syn}
	Y_j= z(t_j;\theta_0) + cU_j.
\end{equation}
Here, $U_j$ is a standard normal random variable ($U_j\sim\mathcal{N}(0,1)$) 
and the constant $c$ is the product of a pre-selected value, $\alpha$,
and the minimum value of the simulated incidence:
\begin{equation}\label{constvar}
	c=\alpha \left[\min_{1\leq i\leq n}z(t_i;\theta_0)\right].
\end{equation}
The multiplier $\alpha$ allows us to control the variance of the noise, while
the use of the minimum incidence is an attempt to reduce the occurrence of
negative values in the synthetic data set.
It is clear from Equation (\ref{noise_syn}) that the noise added to 
the synthetic data has constant variance, given by $var(cU_j)=c^2$. 
A realization of the case counting process is denoted by 
$\{y_j\}_{j=1}^n$ with $y_j=z(t_j;\theta_0) + cu_j$, where all the $u_j$'s 
are independently drawn from a standard normal distribution.

The optimization routine requires an initial estimated value of the parameter;
this is taken to be $\theta=(1+a)\theta_0$, where $a$ also denotes a 
pre-selected multiplier.
Selecting different values for $\alpha$ and $a$ allows us to investigate
the performance of the estimation process in the face of different levels
of noise and differing levels of information as to the approximate location
of the best fitting model parameter (i.e., the ``true'' parameter).

\begin{figure}
\begin{center}
\includegraphics[width=6.5in]{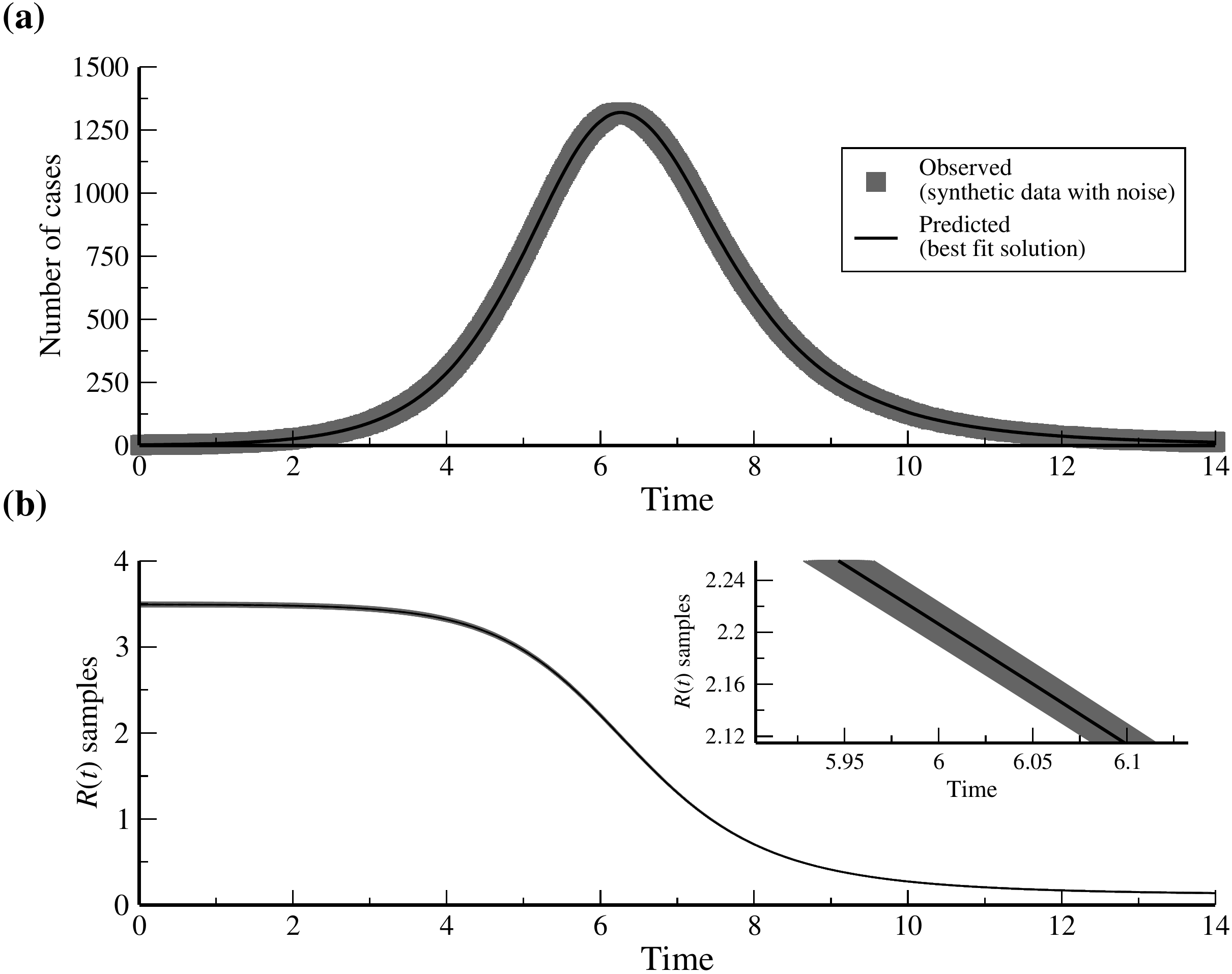}	
\end{center}
\caption{Results obtained by applying the OLS methodology to synthetic data 
with $n=1,000$ observations.  Panel (a) depicts the best fit solution 
(solid curve), and the synthetic data with noise (solid squares), respectively.  Panel (b) displays $1,000$ of the $m=10,000$ effective reproductive number curves (solid gray) constructed using parameters drawn from the 4-multivariate
 normal distribution $\mathcal{N}_4(\hat\theta_{OLS},\hat\Sigma^n_{OLS})$.  The curve $\mathcal{R}(t;\hat\theta_{OLS})$ is shown in 
solid black.  The inset depicts a close-up view of the curves for $t$ in a
small interval about $t=6.0$.}
\label{sim_d_ztrt}
\end{figure}

A synthetic data set with $n=1,000$ observations was constructed by 
setting $\alpha=0.50$.  The initial guess was set using $a=0.25$. Then
$m=10,000$ sample trajectories of $\mathcal{R}(t)$ were generated using
the procedure discussed above. The
resulting estimates of the parameters and effective reproductive numbers,
together with measures of uncertainty, are given in Table 
\ref{table_ols_syntd}. Also listed are the initial parameter guesses given
to the optimization routine and the minimized value of the cost function,
$J(\hat\theta_{OLS})$.   Figure \ref{sim_d_ztrt}(a) depicts the synthetic 
data (squares), together with the best fitting model (solid curve). 
We remark that the observation noise,
which is on the order of $\alpha=0.50$ times the smallest incidence
value, represents a very small error over the major part of the synthetic
data set. As such, it is almost impossible to distinguish between the
data and fitted model in this figure. 

The trajectories of the effective
reproductive number are shown as grey solid curves in Panels (a) and (b)
of Figure \ref{sim_d_ztrt}, in
which the trajectory $\mathcal{R}(t;\hat\theta_{OLS})$ appears as a solid 
black curve. Again, the small errors make it difficult to distinguish between
the central trajectory and the ensemble of  $\mathcal{R}(t)$
trajectories.

\begin{figure}
\begin{center}
\includegraphics[width=6in]{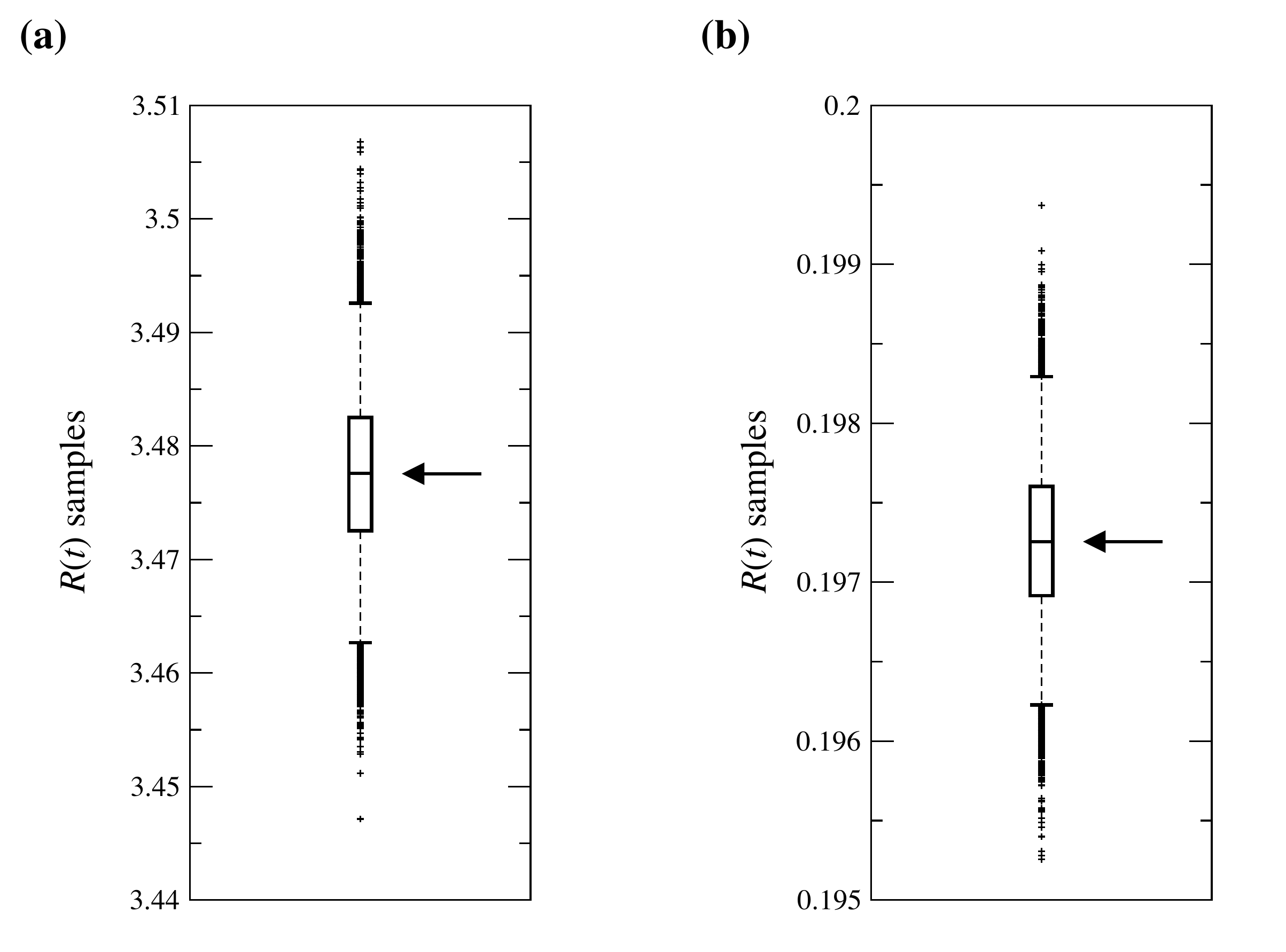}	
\end{center}
\caption{Variability in the samples of $\mathcal{R}(t)$ for two fixed values 
of $t$: (a) t=$2.1$ and (b) $t=11.2$. The box plots depict the 25, 50 and 
75 percentiles of the distribution of $m=10,000$ $\mathcal{R}(t)$ 
samples (lower edge, middle and upper edge, respectively, of the solid box), 
together with the 2.5 and 97.5 percentiles (lower and upper whiskers). 
Samples in the lower and upper 2.5 percentiles are shown as crosses.
Arrows depict the locations of the corresponding central estimates
$\mathcal{R}(t;\hat\theta_{OLS})$.}
\label{boxplotsrt}
\end{figure}

Figure \ref{boxplotsrt} contains box plots of the $\mathcal{R}(t)$ samples at
two fixed times: (a) $t=2.01$, and (b) $t=11.2$.  We use the
2.5 and 97.5 percentiles of the $\mathcal{R}(t)$ sample distribution at
time $t$ to quantify uncertainty in the central estimate of $\mathcal{R}(t)$. 
At the bottom of Table \ref{table_ols_syntd} the estimates of the effective
reproductive number are summarized by showing the minimum and maximum 
(over time) of the central estimate of $\mathcal{R}(t)$ together with the 
uncertainty bounds obtained at these two time points.

\begin{table}
\centering
\caption{Estimates obtained using synthetic data with constant variance 
noise ($\alpha=0.50$, see text for further details).  The number of 
observations is $n=1,000$, while the $\mathcal{R}(t)$ sample size is 
$m=10,000$. The optimization algorithm was initialized with the parameter
value $\theta=1.25\theta_0$, where $\theta_0$ 
denotes the true parameter.  The true value, initial guess, estimate,
and standard error, are given for all parameters, along with the value of 
the cost function evaluated at the parameter estimate. The
minimum (Min.) and maximum (Max.) of the central estimate of the
effective reproductive number ($\mathcal{R}(t;\hat\theta_{OLS})$) are given 
with the accompanying 2.5 and 97.5 percentiles (in square brackets).}

\begin{tabular}{|c|c|c|c|c|c|} \hline\hline
Parameter   &   True value   &   Initial guess       &    Estimate    &    Standard error\\ \hline
$S_0$&3.500$\times 10^{5}$&4.375$\times 10^{5}$&3.501$\times 10^{5}$&1.065$\times 10^{2}$\\ \hline
$I_0$&9.000$\times 10^{1}$&1.125$\times 10^{2}$&8.987$\times 10^{1}$&1.966$\times 10^{-1}$\\ \hline
$\tilde\beta$&5.000$\times 10^{-6}$&6.250$\times 10^{-6}$&5.003$\times 10^{-6}$&$3.794\times 10^{-9}$\\ \hline
$\gamma$&5.000$\times 10^{-1}$&6.250$\times 10^{-1}$&5.013$\times 10^{-1}$&1.609$\times 10^{-3}$\\ \hline\hline
\multicolumn{5}{|c|}{$J(\hat\theta_{OLS})=1.099 \times 10^{3}$}\\
\multicolumn{2}{|c}{$\sigma_0^2=1.237 \times 10^{0}$}
&\multicolumn{1}{c}{}
&\multicolumn{2}{c|}{$\hat\sigma^2_{OLS}=1.104  \times 10^{0}$} \\ \hline\hline
\multicolumn{3}{|c|}{Min. $\mathcal{R}(t;\hat\theta_{OLS})$ }&\multicolumn{2}{c|}{0.138\ \ [0.137,0.140]}\\ \hline
\multicolumn{3}{|c|}{Max. $\mathcal{R}(t;\hat\theta_{OLS})$ }&\multicolumn{2}{c|}{3.494\ \ [3.478,3.509]}\\ \hline
\end{tabular}

\label{table_ols_syntd}
\end{table}

\begin{figure}
\begin{center}
\includegraphics[width=6in]{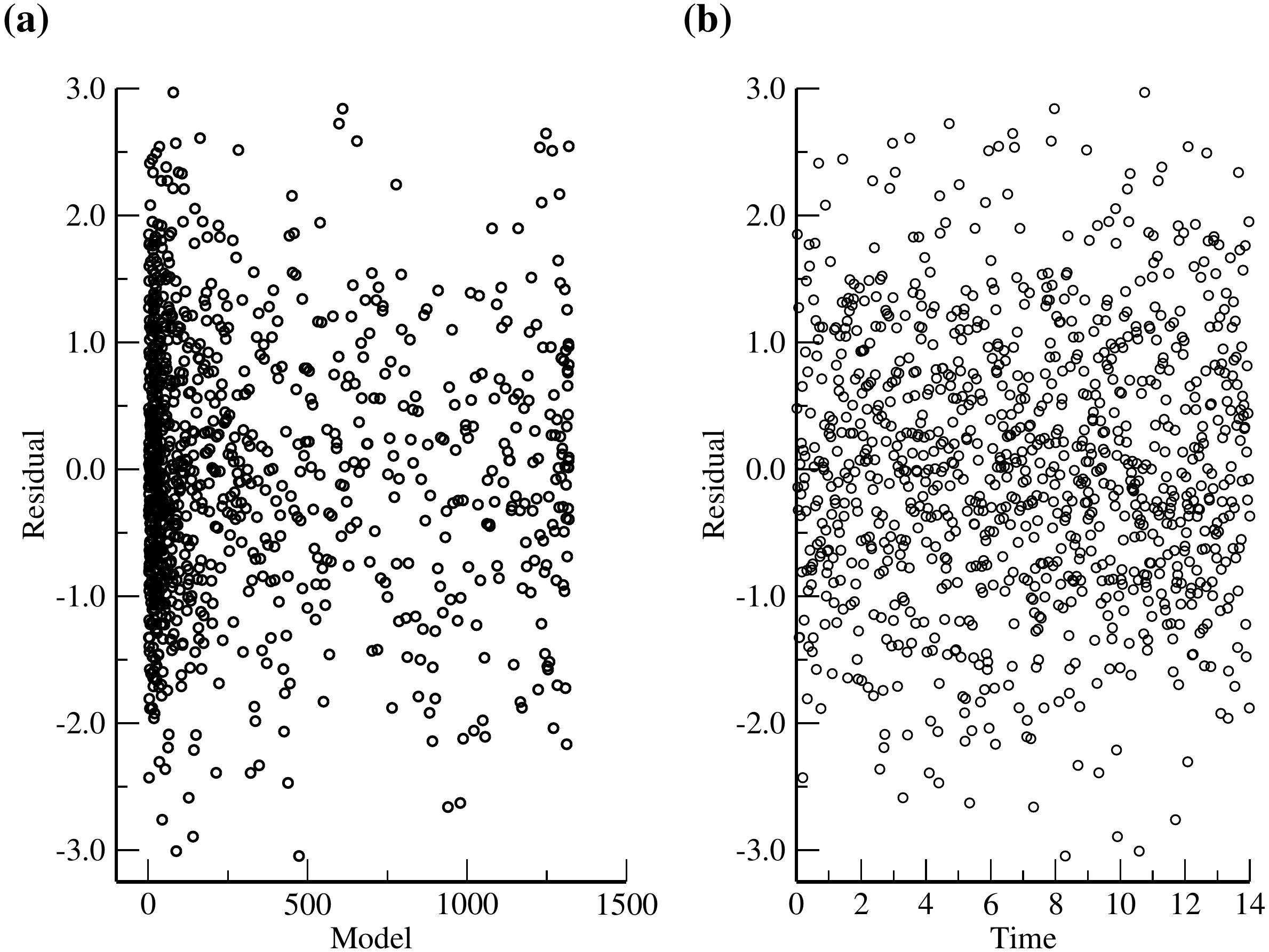}	
\end{center}
\caption{Analysis of the residuals from the OLS estimation applied to
the synthetic data. Panel (a) depicts the residuals 
$y_j-z(t_j;\hat\theta_{OLS})$ versus the model values 
$z(t_j;\hat\theta_{OLS})$ for $j=1,\dots,n$.  Panel (b) displays the 
residuals versus time $t_j$ for $j=1,\dots,n$.}
\label{sim_d_ztrt_res}
\end{figure}

A simple residuals analysis is illustrated in Figure \ref{sim_d_ztrt_res}.  
A residual at time $t_j$ is defined as $y_j-z(t_j;\hat\theta_{OLS})$.  
In Figure \ref{sim_d_ztrt_res}(a), these residuals are plotted against the
predicted values, $z(t_j;\hat\theta_{OLS})$.  Figure \ref{sim_d_ztrt_res}(b) 
displays a plot of the residuals against time.  No patterns or trends are
seen in these residual plots (for example, the magnitudes of the residuals
show no trends in either plot and the residuals do not exhibit any temporal
patterns or correlations). This is to be expected because the
$u_j$ are realizations of independent (uncorrelated) and identically 
distributed (standard normal) random variables.

In this example, the OLS methodology performs well, yielding excellent
estimates of the true parameter value. This should not be surprising because
the noise level was chosen to be extremely small and we provided the 
optimization routine with an initial parameter value that was close to
the true value.

The application of the OLS methodology does not always go so smoothly
as in the previous example: parameter estimates can be obtained that are
far from the true values. The cost function, $J(\theta)$, typically
possesses multiple minima and the simple-minded use of {\tt fminsearch}
can yield a parameter estimate located at one of the other (local) 
minima, particularly when the initial parameter estimate is some distance
away from the true value.

Table \ref{tab_mult_min} presents estimates for the same synthetic data
set, but for which the initial parameter estimate was taken to be
one hundred and seventy five percent ($\theta=2.75\theta_0$) away from 
the true parameter value. This results in poor estimates of the parameters:
the values of $S_0$, $\tilde\beta$ and $\gamma$ are overestimated by 317, 
42, and 1730 percent, respectively, while the value of $I_0$ is 
underestimated by 78 percent. Worryingly, the values of the standard
errors give no warning that the parameter estimates are quite so poor:
the largest standard error, relative to the parameter estimate, is
obtained for $\gamma$ and equals 14\%, while these figures fall to 12\%,
10\% and less than 1\% for the remaining parameters.

\begin{figure}
\begin{center}
\includegraphics[width=6in]{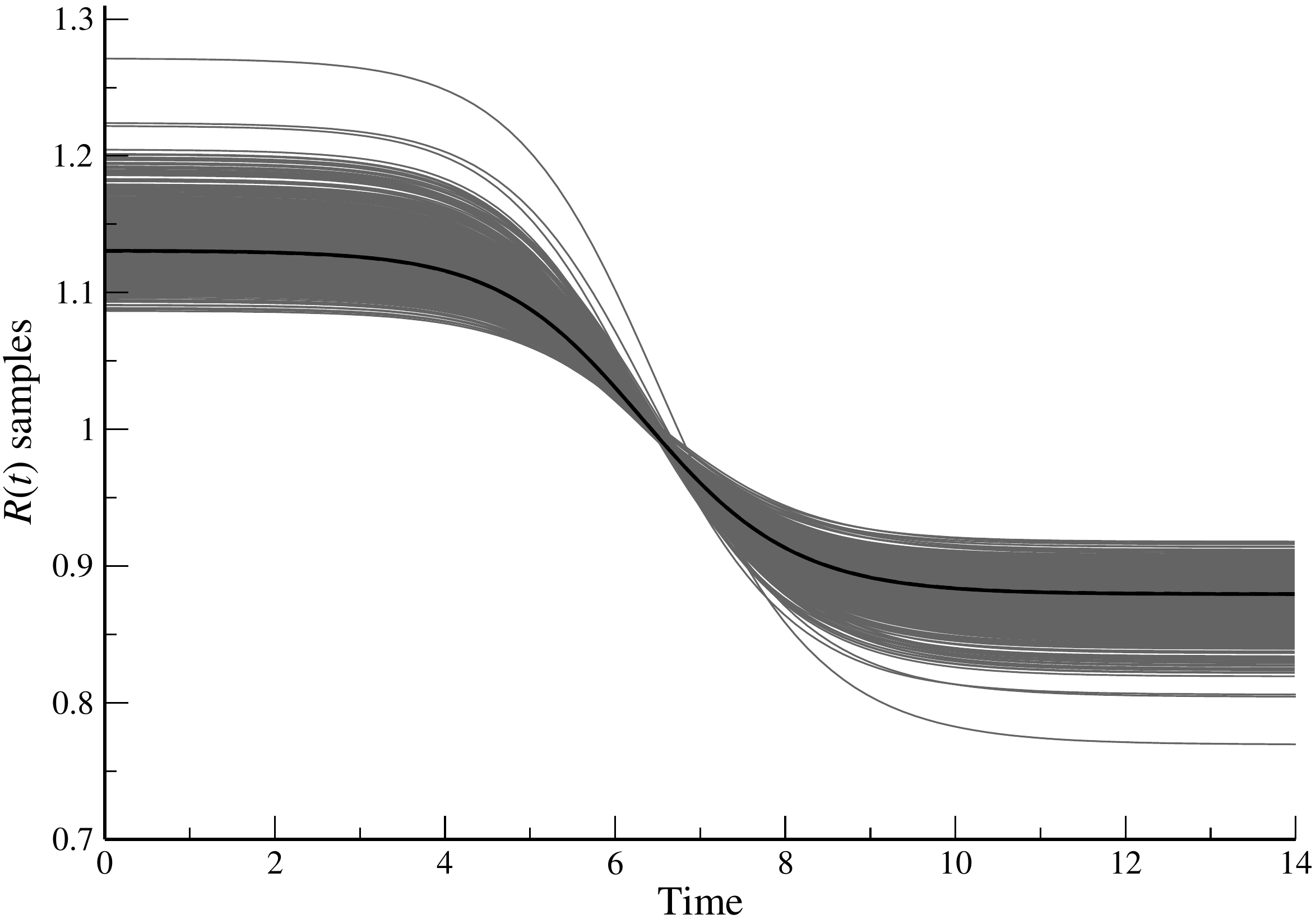}
\end{center}
\caption{One thousand of the $m=10,000$ effective reproductive number 
curves (solid gray) constructed using parameters drawn from the 4-multivariate
 normal distribution $\mathcal{N}_4(\hat\theta_{OLS},\hat\Sigma^n_{OLS})$.  
The curve $\mathcal{R}(t;\hat\theta_{OLS})$ is shown in solid black. The
curve of the median value of the $\mathcal{R}(t)$ samples, at each $t$,
is also shown as a dashed black curve, but is indistinguishable from
the curve $\mathcal{R}(t;\hat\theta_{OLS})$.}
\label{ols_rt_syntd_min2_mean_med}
\end{figure}

The true maximum value of $\mathcal{R}(t)$ 
is 3.500, yet the effective reproductive number has an estimated upper bound 
of 1.131; clearly the misleading estimates of $S_0$ and $\gamma$
cause $\mathcal{R}(t)$ to be underestimated.  If we did not know 
the true value of the effective reproductive number, we would be unlikely 
to anticipate this underestimation, because the estimate and percentiles 
in this case,  $[1.102,1.182]$, do not suggest that there is a large 
uncertainty. However the issues with the estimation of the individual
parameters alert us to possible problems. Interestingly, the 
distribution of $\mathcal{R}(t)$ samples are no longer
normally distributed about the central estimate (Figure 
\ref{ols_rt_syntd_min2_mean_med}).
	 
\begin{figure}
\begin{center}
\includegraphics[width=6in]{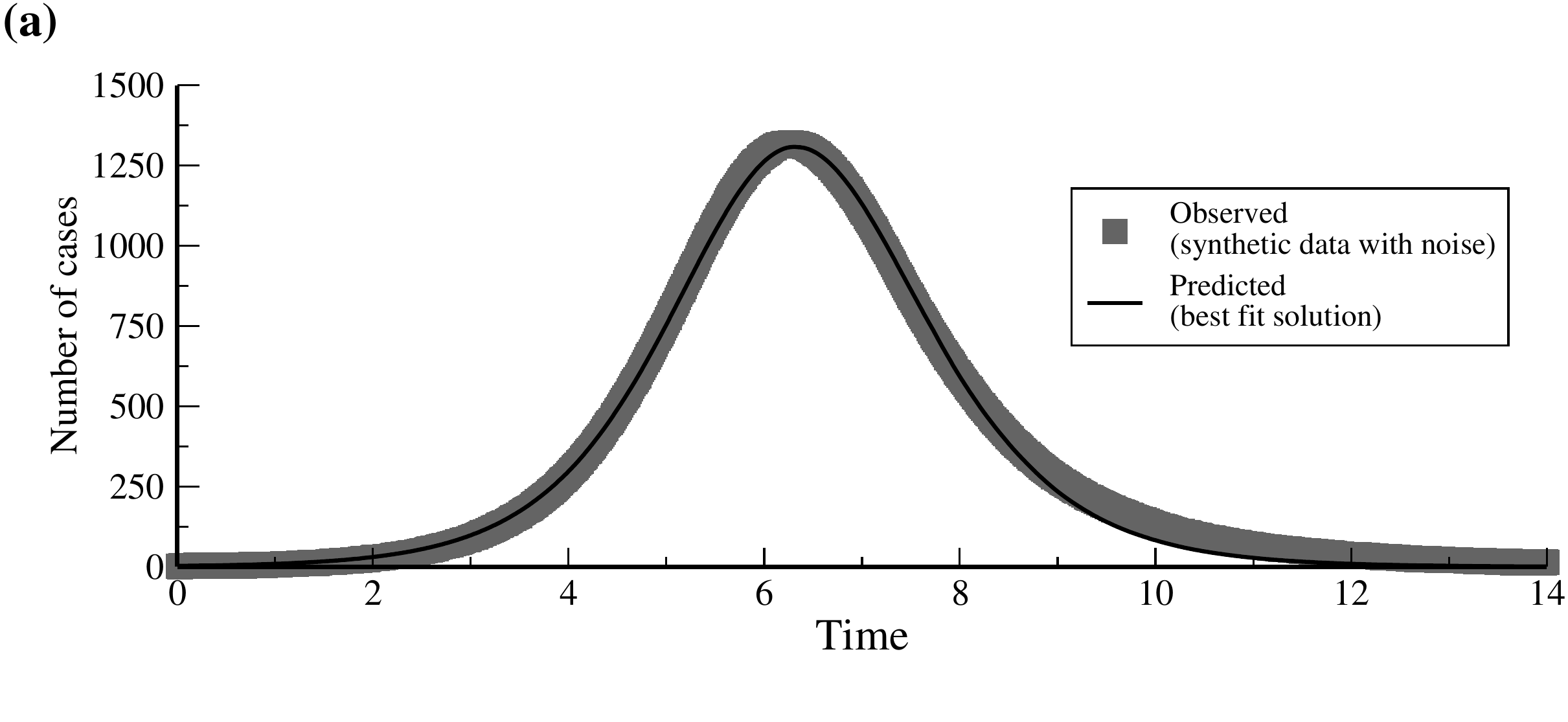}
\includegraphics[width=6in]{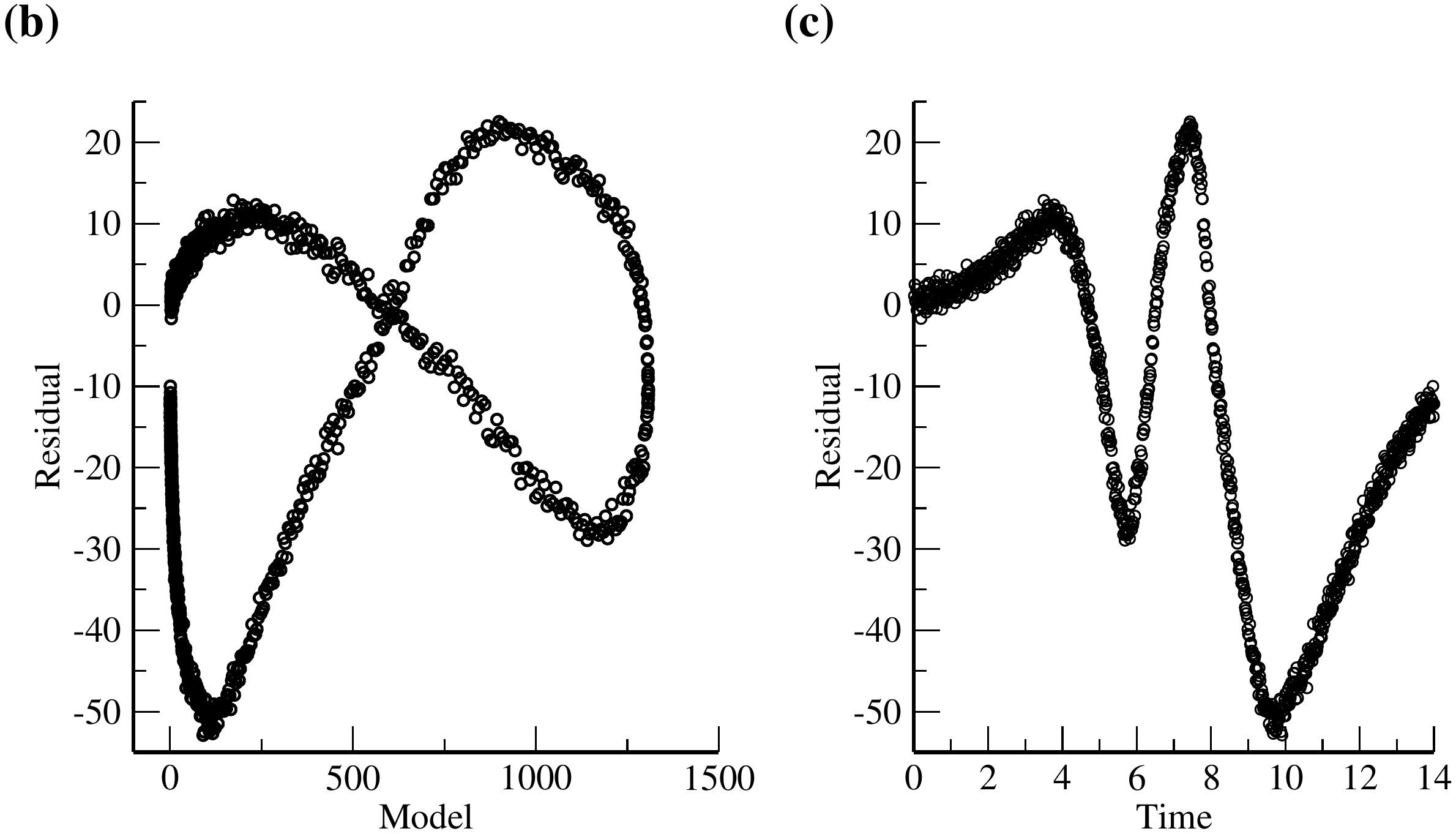}	
\end{center}
\caption{Results obtained using $\theta=2.75\theta_0$ as the initial guess in the optimization algorithm, applying the OLS methodology to synthetic data 
with $n=1,000$ and $\alpha=0.50$. 
Panel (a) displays the model prediction (solid curve), and the observations (solid squares), respectively.  
Panel (b) displays the residuals against the model values and panel (c)
displays the residuals versus time.}
\label{bad_fit}
\end{figure}

Because we know the true parameter value and the outcome of a successful
model fit to this data set, it was easy for us to identify the problems
that arose here. We note, for instance, that the value of the cost function 
for the estimated parameter value is two orders of magnitude larger than 
in the previous case, quantifying that the model fit is much worse. We
would not have the luxury of these pieces of information if this
estimation arose in the consideration of a real-world data set. The
residuals plots, however, clearly suggest that there are serious problems with
the model fit. In particular, there are obvious temporal trends in the 
residuals, indicating systematic deviations between the fitted model and 
data. Even though the observation noise is small, it is just possible to 
see these deviations in Figure \ref{bad_fit}(a), but they are considerably 
easier to spot in the residuals plots of Figures \ref{bad_fit}(b) and (c).

\begin{table}
\centering
\caption{Parameter estimates from synthetic data ($n=1,000$ observations) with 
constant variance noise
($\alpha=0.50$) using $\theta=2.75\theta_0$ as the initial guess in the
optimization algorithm.  The sample size for $\mathcal{R}(t)$ is $m=10,000$.}
\begin{tabular}{|c|c|c|c|c|c|} \hline\hline
Parameter   &   True value   &   Initial guess       &    Estimate    &    Standard error\\ \hline
$S_0$&3.500$\times 10^{5}$&9.625$\times 10^{5}$&1.459$\times 10^{6}$&1.772$\times 10^{5}$\\ \hline
$I_0$&9.000$\times 10^{1}$&2.475$\times 10^{2}$&1.957$\times 10^{1}$&1.913$\times 10^{0}$\\ \hline
$\tilde\beta$&5.000$\times 10^{-6}$&1.375$\times 10^{-5}$&7.098$\times 10^{-6}$&2.662$\times 10^{-8}$\\ \hline
$\gamma$&5.000$\times 10^{-1}$&1.375$\times 10^{0}$&9.160$\times 10^{0}$&1.274$\times 10^{0}$\\ \hline\hline
\multicolumn{5}{|c|}{$J(\hat\theta_{OLS})=5.631 \times 10^{5}$}\\
\multicolumn{2}{|c}{$\sigma_0^2=1.237  \times 10^{0}$}
&\multicolumn{1}{c}{}
&\multicolumn{2}{c|}{$\hat\sigma^2_{OLS}=5.653 \times 10^{2}$} \\ \hline\hline
\multicolumn{3}{|c|}{Min. $\mathcal{R}(t;\hat\theta_{OLS})$}&\multicolumn{2}{c|}{0.879\ \ [0.837,0.904]}\\ \hline
\multicolumn{3}{|c|}{Max. $\mathcal{R}(t;\hat\theta_{OLS})$}&\multicolumn{2}{c|}{1.131\ \ [1.102,1.182]}\\ \hline
\end{tabular}
\label{tab_mult_min}
\end{table}

\clearpage

\subsection{Synthetic Data with Non-constant Variance Noise}

We generated a second synthetic data set with non-constant variance noise. 
The true value $\theta_0$ was fixed, and was used to calculate the 
numerical solution $z(t_j;\theta_0)$.  Observations were computed 
in the following fashion:  
\begin{equation}
\label{GLS_stat_model}
	Y_j=z(t_j;\theta_0)\left(1+\alpha V_j\right),
\end{equation}
where the $V_j$ are independent random variables with standard normal 
distribution, and $0<\alpha<1$ denotes a desired percentage.  
In this way, $var(Y_j)=[z(t_j;\theta_0)\alpha]^2$ which is non-constant 
across the time points $t_j$.  If the terms $\{v_j\}_{j=1}^n$ denote a 
realization of $\{V_j\}_{j=1}^n$, then a realization of the observation 
process is denoted by $y_j=z(t_j;\theta_0)(1+\alpha v_j)$.

An $n=1,000$ point synthetic data set was constructed with $\alpha=0.075$.  
The optimization algorithm was initialized with the estimate 
$\theta=1.10\theta_0$.  The weights in the normal equations defined 
by Equation (\ref{glsdef}), were chosen as $w_j=1/z(t_j;\theta)^{2}$. 

\begin{figure}
\centering
\includegraphics[width=5in]{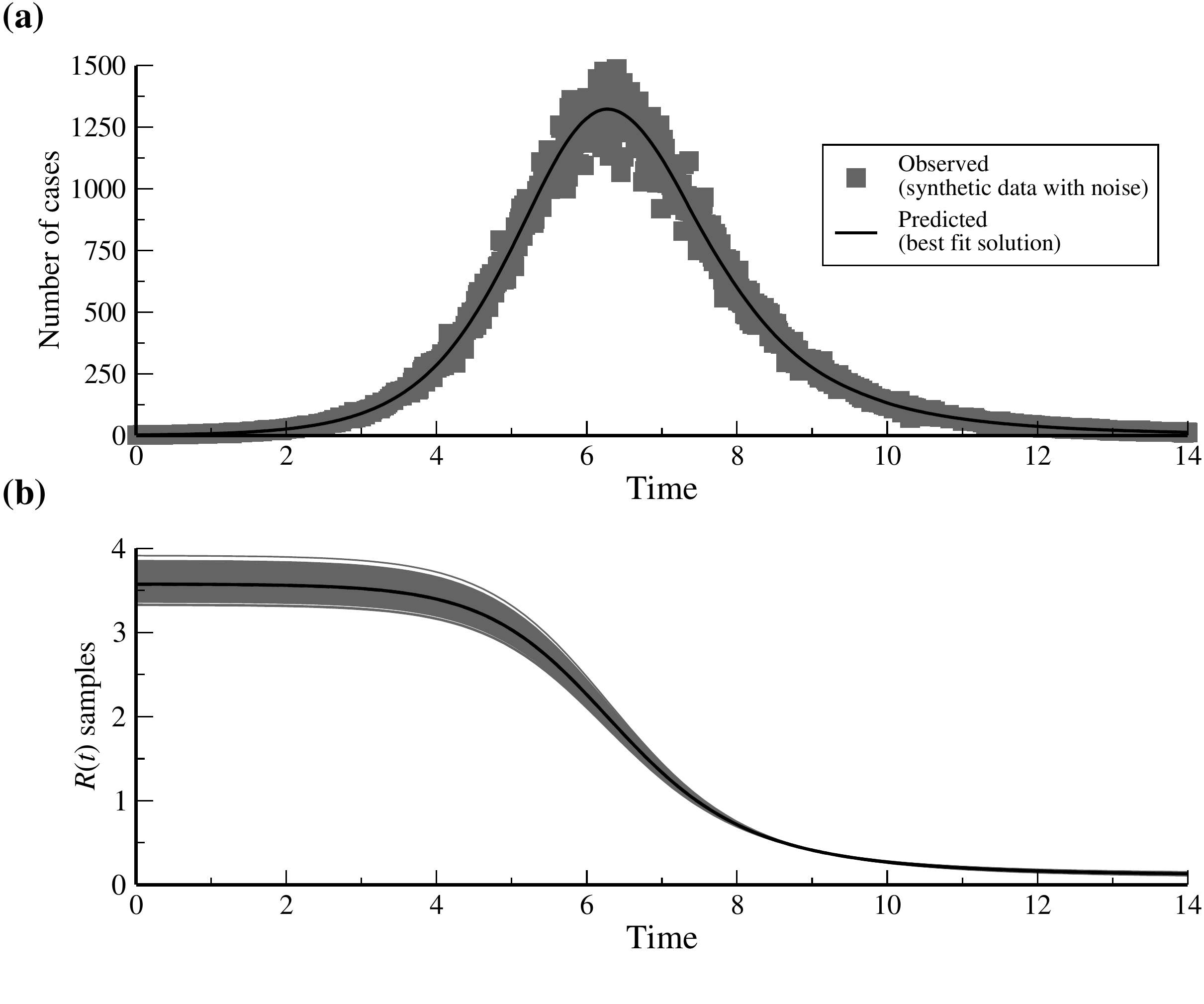}
\includegraphics[width=5in]{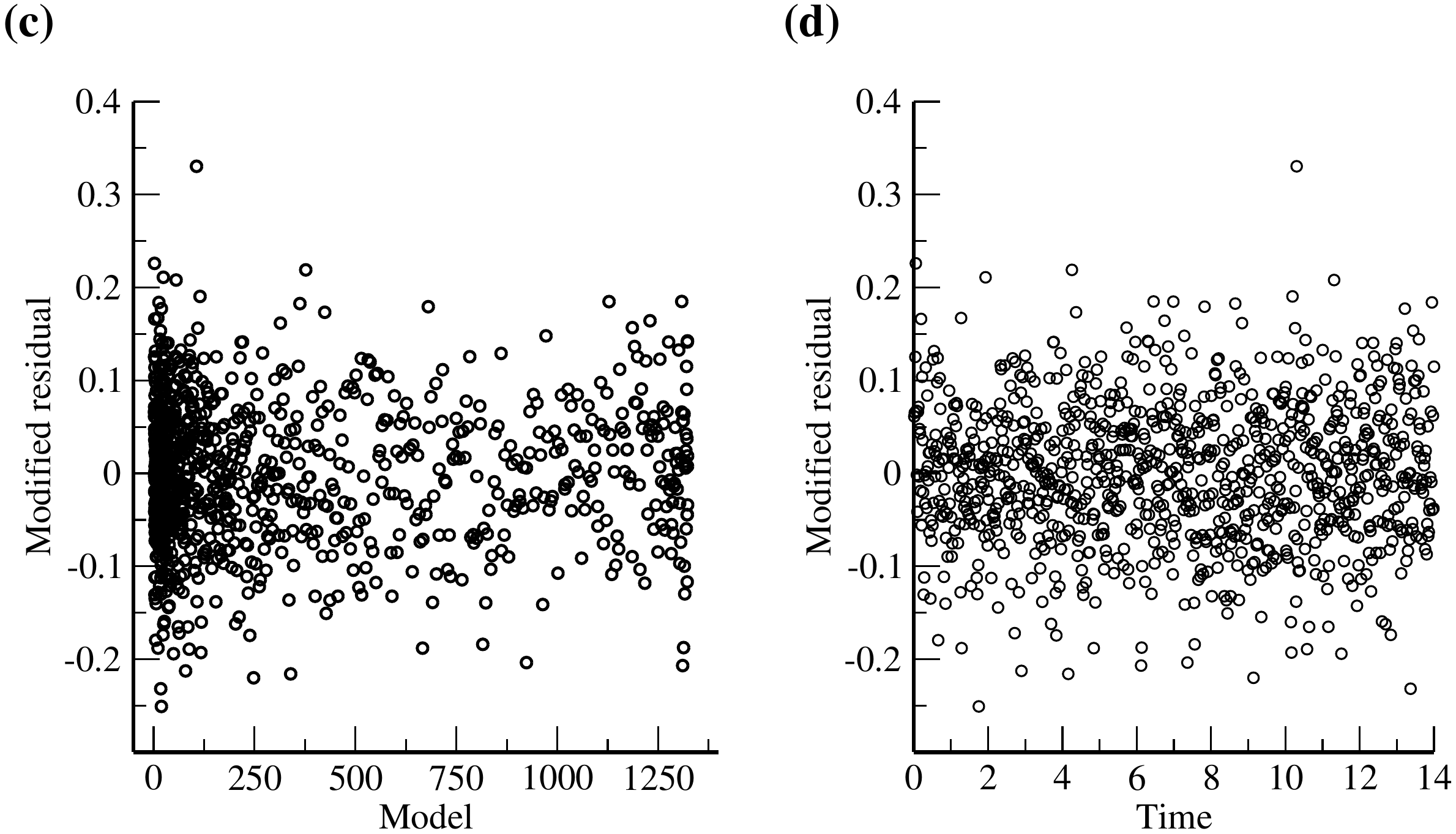}
\caption{Results from applying the GLS methodology to synthetic data 
with non-constant variance noise ($\alpha=0.075$), using $n=1,000$ 
observations.  The initial guess for the optimization routine
was $\theta=1.10\theta_0$.  The weights in the cost function were equal 
to $1/z(t_j;\theta)^{2}$, for $j=1,\dots,n$.  Panel (a) depicts the observed 
and fitted values and panel (b) displays $1,000$ of the $m=10,000$ 
$\mathcal{R}(t)$ sample trajectories. Residuals plots are presented in
panels (c) and (d): modified residuals versus fitted values in (c) and 
modified residuals versus time in (d).}    
\label{gls_syntd_fig}
\end{figure}

Table \ref{gls_syntd_table} lists estimates of the parameters and 
$\mathcal{R}(t)$, together with uncertainty estimates. In the case of
$\mathcal{R}(t)$, uncertainty was assessed based on the simulation approach
using $m=10,000$ samples of the parameter vector, drawn from
 $\mathcal{N}_4(\hat\theta_{GLS},\hat\Sigma^n_{GLS})$.
Figure \ref{gls_syntd_fig}(a) depicts both data and fitted model points,
 $z(t_j;\hat\theta_{GLS})$, plotted versus $t_j$. Figure 
\ref{gls_syntd_fig}(b) depicts $1,000$ of the $10,000$ $\mathcal{R}(t)$ curves.

\begin{table}[h]
\centering
\caption{Estimates from a synthetic data of size $n=1,000$, with non-constant 
variance using  $\alpha=0.075$.  The 
$\mathcal{R}(t)$ sample size is $m=10,000$.  The initial guess of the
optimization algorithm was $\theta=1.10\theta_0$.  Each weight 
in the cost function $L(\theta)$ (see Equation (\ref{costgls})) was equal 
to $1/z(t_j;\theta)^{2}$ for $j=1,\dots,n$.}
\begin{tabular}{|c|c|c|c|c|c|} \hline\hline
Parameter   &   True value   &   Initial guess       &    Estimate    &    Standard error\\ \hline
$S_0$&3.5000$\times 10^{5}$&3.800$\times 10^{5}$&3.498$\times 10^{5}$&1.375$\times 10^{3}$\\ \hline
$I_0$&9.000$\times 10^{1}$&9.900$\times 10^{1}$&9.085$\times 10^{1}$&1.424$\times 10^{0}$\\ \hline
$\tilde\beta$&5.000$\times 10^{-6}$&5.500$\times 10^{-6}$&4.954$\times 10^{-6}$&4.411$\times 10^{-8}$\\ \hline
$\gamma$&5.000$\times 10^{-1}$&5.500$\times 10^{-1}$&4.847$\times 10^{-1}$&1.636$\times 10^{-2}$\\ \hline
\multicolumn{5}{|c|}{$L(\hat\theta_{GLS})=5.689\times 10^{0}$}\\ 
\multicolumn{2}{|c}{$\sigma_0^2=5.625 \times 10^{-3}$}
&\multicolumn{1}{c}{}
&\multicolumn{2}{c|}{$\hat\sigma^2_{GLS}=5.712 \times 10^{-3}$} \\ \hline\hline
\multicolumn{3}{|c|}{Min. $\mathcal{R}(t;\hat\theta_{GLS})$}&\multicolumn{2}{c|}{0.132\ \ [0.120,0.146]}\\ \hline
\multicolumn{3}{|c|}{Max. $\mathcal{R}(t;\hat\theta_{GLS})$}&\multicolumn{2}{c|}{3.576\ \ [3.420,3.753]}\\ \hline
\end{tabular}
\label{gls_syntd_table}
\end{table}

\clearpage

Residuals plots are displayed in Figures \ref{gls_syntd_fig}(c) and (d).  
Because $\alpha v_j=(y_j-z(t_j;\theta_0))/z(t_j;\theta_0)$,
by construction of the synthetic data, the residuals analysis focuses on the 
ratios
\[
	\frac{y_j-z(t_j;\hat\theta_{GLS})}{z(t_j;\hat\theta_{GLS})}
\]
which in the labels of Figures \ref{gls_syntd_fig}(c) and (d) are referred 
to as ``Modified residuals''.  In Figure \ref{gls_syntd_fig}(c)
these ratios are plotted against $z(t_j;\hat\theta_{GLS})$, 
while Panel (d) displays them versus the time points $t_j$.  
The lack of any discernable patterns or trends in 
Figure \ref{gls_syntd_fig}(c) and (d)
confirms that the errors in the synthetic data set conform to the assumptions
made in the formulation of the statistical model of Equation 
(\ref{GLS_stat_model}). In particular, the errors are uncorrelated and
have variance that scales according to the relationship stated above.

\clearpage

\section{Analysis of Influenza Outbreak Data}\label{rslts_ols}

The OLS and GLS methodologies were applied to longitudinal observations
of six influenza outbreaks (see Section \ref{data}), giving	
estimates of the parameters and the reproductive number for each season.
The number of observations $n$ varies from season to season.  
The $\mathcal{R}(t)$ sample size was $m=10,000$ in each case.
The set of admissible parameters $\Theta$ is defined by the lower and 
upper bounds listed in Table \ref{tabbaselinep} along with the inequality 
constraint $S_0\tilde\beta/\gamma>1$. The bounds in Table \ref{tabbaselinep}
were obtained or based on \cite{ccvvb,lkmf,ncwc} and references therein.
For brevity, we only present here the results obtained using data from 
the 1989-1999 season.

\renewcommand{\arraystretch}{1.5}
\begin{table}[h]
\caption{Lower and upper bounds on the initial conditions and parameters.}
\begin{center}
\begin{tabular}{|c|c|}\hline
 Suitable Range & Unit \\ \hline\hline
1.00$\times 10^{2}$\ $ < S_0 < $\ 7.00$\times 10^{6}$ &  people   \\ \hline
0.00\ $< I_0< $\ 5.00$\times 10^{3}$ &   people  \\ \hline
 7.00$\times 10^{-9}$\ $ <\tilde\beta<$\ 7.00$\times 10^{-1}$ &  weeks$^{-1}$people$^{-1}$  \\ \hline
$3/7<1/\gamma<$4/7& weeks\\ \hline
\end{tabular}
\end{center}
\label{tabbaselinep}
\end{table}
\renewcommand{\arraystretch}{1.0}

\subsection{OLS Estimation}\label{olscd}

\begin{figure}
\centering
\includegraphics[width=5in]{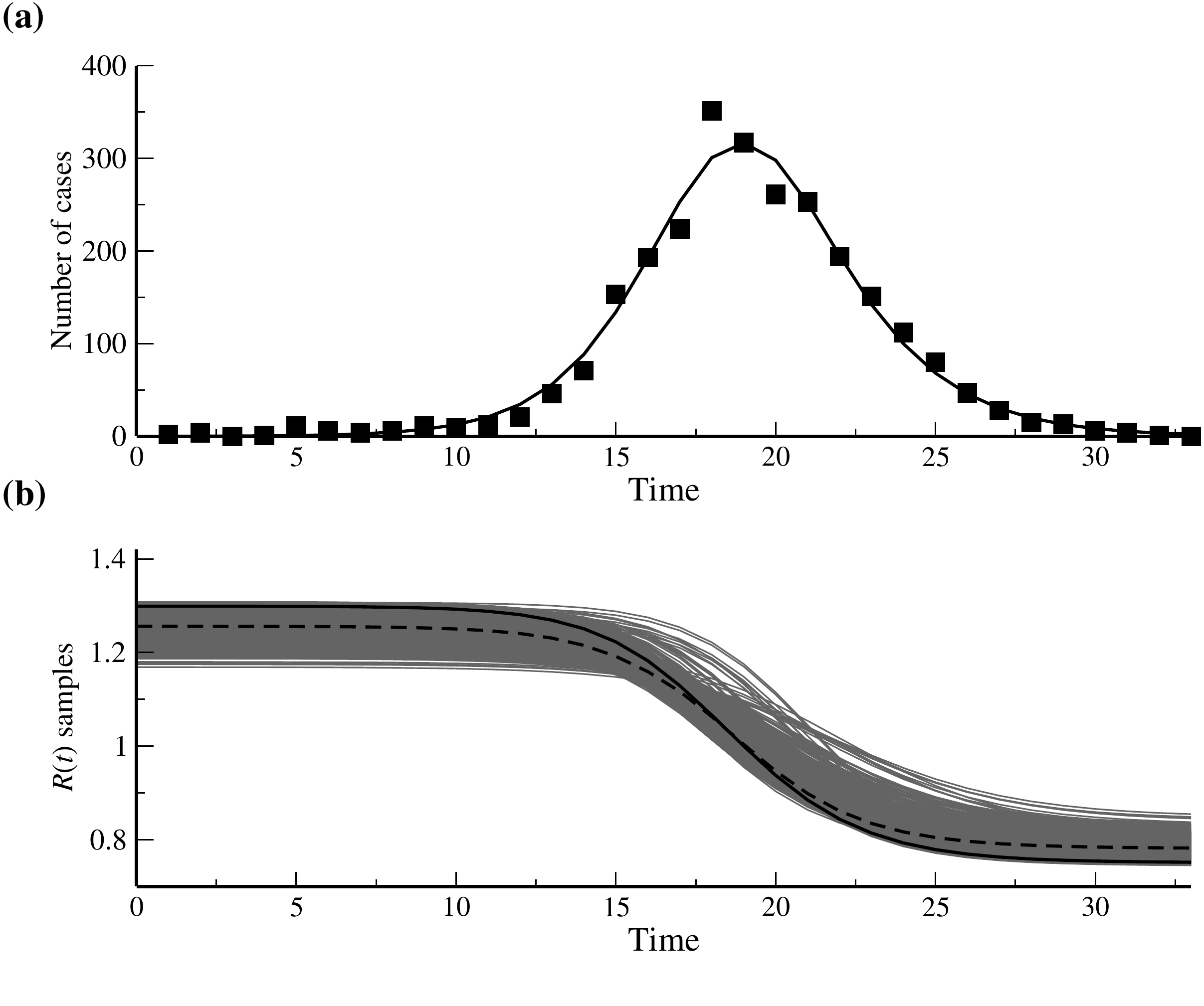}
\includegraphics[width=5in]{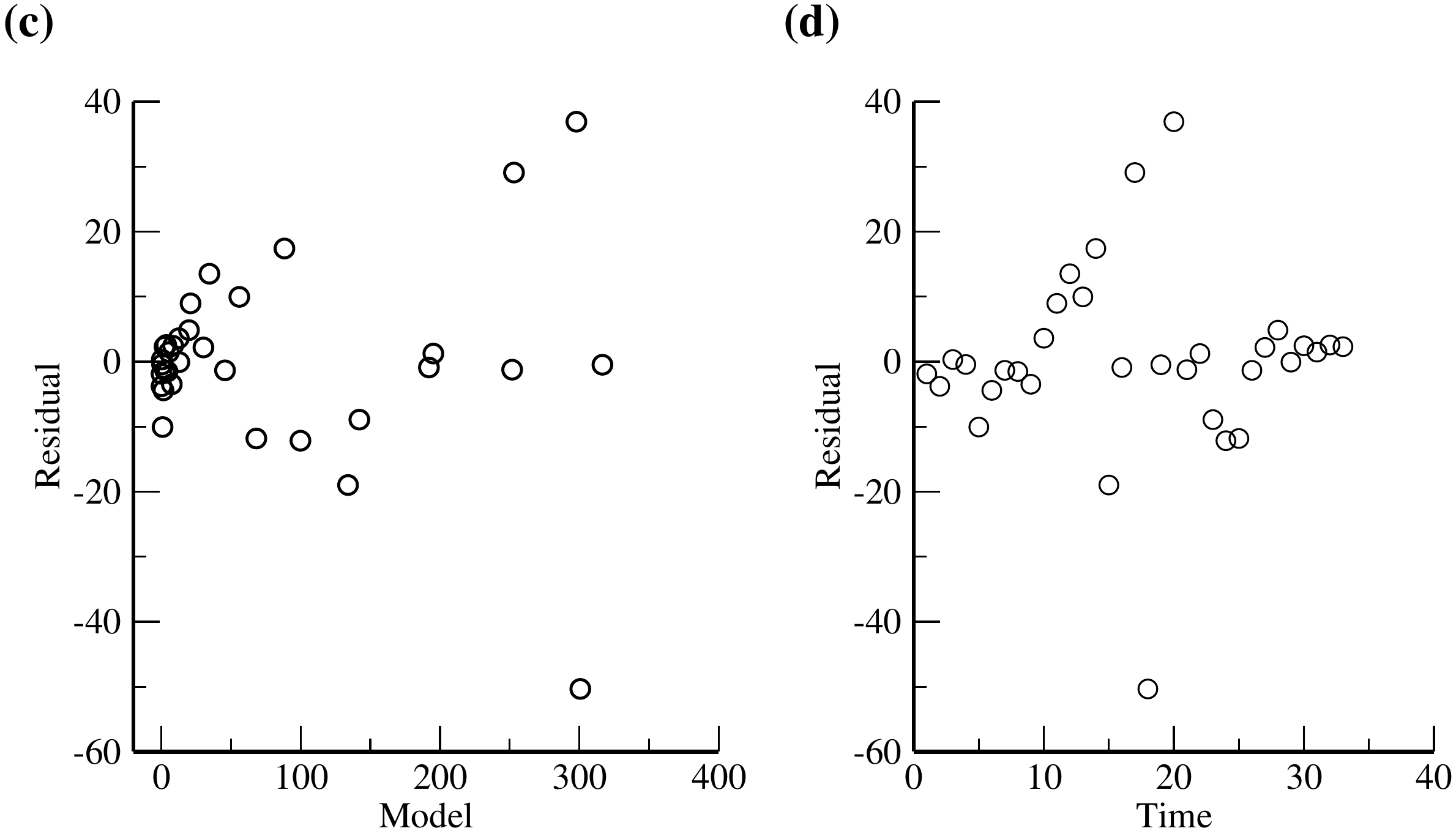}
\caption{OLS model fits to influenza data from season 1998-1999.  
Panel (a) depicts the observations (solid squares) and
the model prediction (solid curve), respectively.  In Panel (b) the 
samples of the effective reproductive number $\mathcal{R}(t)$ are displayed 
(grey curves) together with the central estimate 
$\mathcal{R}(t;\hat\theta_{OLS})$  (solid black curve). The dashed black
curve depicts the median, at each time point, of the distribution of 
the $\mathcal{R}(t)$ samples.
Panel (c) contains the residuals plotted versus the model prediction. 
In Panel (d) residuals are plotted against time.}
\label{9899_ols}
\end{figure}

In most cases, {\it visual} comparison of the trajectory of the best fitting 
model obtained using OLS and the data points suggests that a good fit has 
been achieved (Figure \ref{9899_ols}(a)). 
The statistics that quantify the uncertainty in the estimated values
of the parameters, however, indicate that this may not always be the case.
In many cases, the standard errors are of the same order of magnitude as
the parameter values themselves, indicating wide error bounds (see
Table \ref{9899_ols_table}) and suggesting a lack of confidence in the
estimates. 

We should,
however, interpret the statistical results with some caution because the 
residuals plots (Figure \ref{9899_ols}(c) and (d)) show clear patterns, 
indicating that the assumptions
of the statistical model may have been violated. For instance, the variance
of the residuals appears to increase with the predicted value. There are
definite patterns visible in the residuals versus time plot. The temporal
correlation of the errors could
represent some inadequacy in the way that the dynamic model describes 
the epidemic process, or an inadequacy in the data set itself.

Another indication of problems in the estimation process comes from
the matrix \linebreak $\chi(\hat\theta_{OLS},n)^T\chi(\hat\theta_{OLS},n)$.
The condition number of this matrix is $9.4\times 10^{19}$, indicating
that the matrix is close to singular. Calculation of the covariance
matrix requires the inversion of this matrix, and, as mentioned above,
the asymptotic theory requires that the matrix $\Omega_0$, defined
as a limit of matrix products of this form, has a non singular limit
as $n\rightarrow\infty$. A nearly singular matrix can arise when
there is redundancy in the data or when there are problems with
parameter identifiability \cite{banksdav,Banksernst}. 

It is interesting to observe that the estimated values of $\gamma$ 
frequently fall on the boundary of the feasible region.
This may impact the uncertainty analysis, given that the conditions of the 
asymptotic theory require that the true parameter value lies in the 
interior of the feasible region. If our estimates commonly fall on 
the boundary, this could be
an indication that the true parameter value may not lie within our
feasible region. 
We return to this issue below, in Section \ref{gls3parcd}.

\begin{table}
\centering
\caption{Season 1998-1999.  Parameter and effective reproductive number 
estimates obtained using OLS.}
\label{9899_ols_table}
\begin{tabular}{|c|c|c|} \hline\hline
Parameter   & Estimate    &    Standard error\\ \hline
$S_0$&6.200$\times 10^{3}$&4.514$\times 10^{3}$\\ \hline
$I_0$&3.878$\times 10^{-2}$&1.812$\times 10^{-2}$\\ \hline
$\tilde\beta$&3.667$\times 10^{-4}$&2.198$\times 10^{-5}$\\ \hline
$\gamma$&1.750$\times 10^{0}$&1.753$\times 10^{0}$\\ \hline
\multicolumn{3}{|c|}{$J(\hat\theta_{OLS})=6.357\times 10^{3}$}\\ 
\multicolumn{3}{|c|}{ $\hat\sigma^2_{OLS}=2.192\times 10^{2}$ }\\ \hline\hline
\multicolumn{1}{|c|}{Min. $\mathcal{R}(t;\hat\theta_{OLS})$}&\multicolumn{2}{c|}{0.752\ \ [0.752,0.824]}\\ \hline
\multicolumn{1}{|c|}{Max. $\mathcal{R}(t;\hat\theta_{OLS})$}&\multicolumn{2}{c|}{1.299\ \ [1.202,1.300]}\\ \hline
\end{tabular}
\end{table}
\clearpage

\subsection{GLS Estimation}\label{glscd}
	
\begin{figure}

\centering
\includegraphics[width=5in]{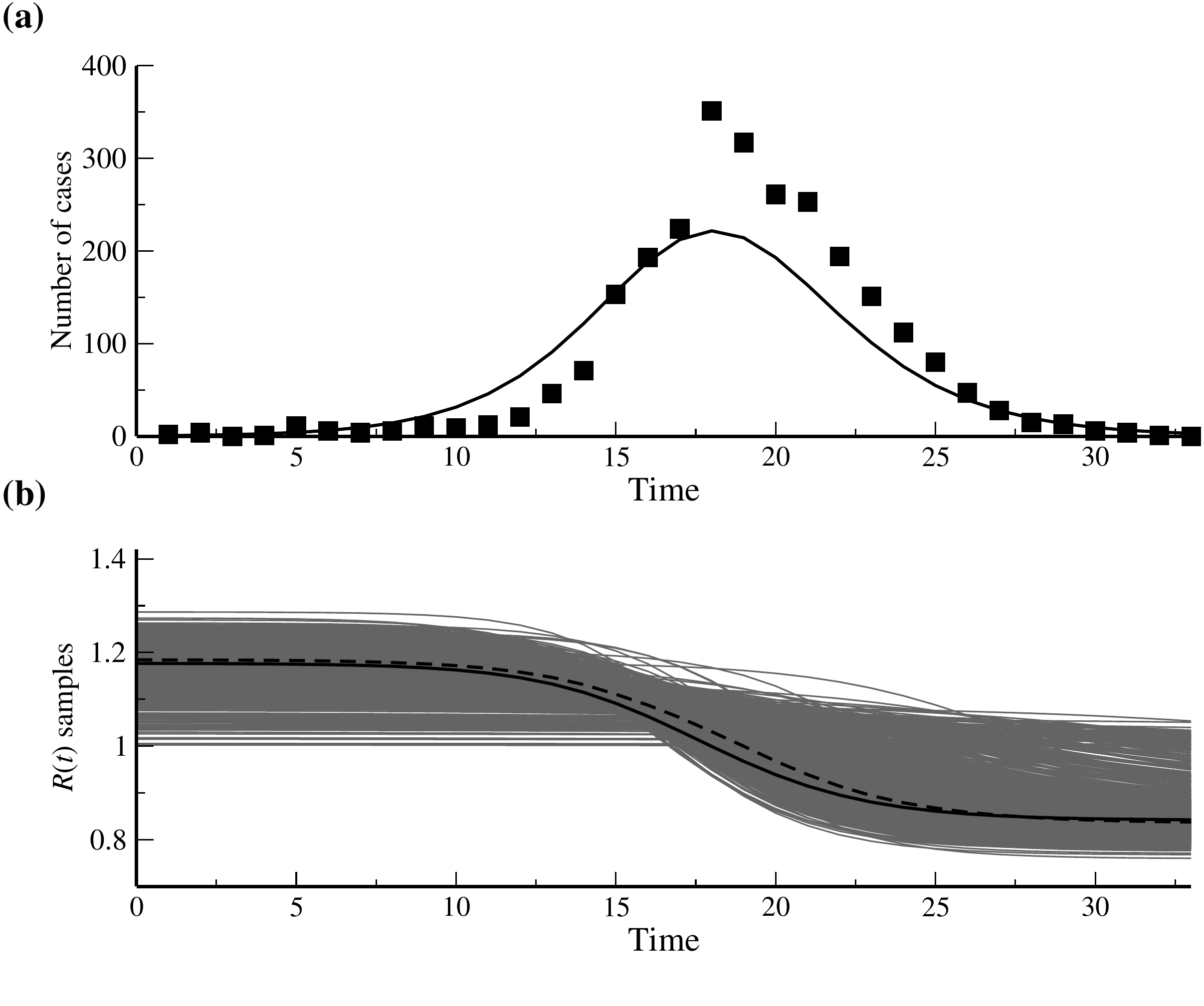}
\includegraphics[width=5in]{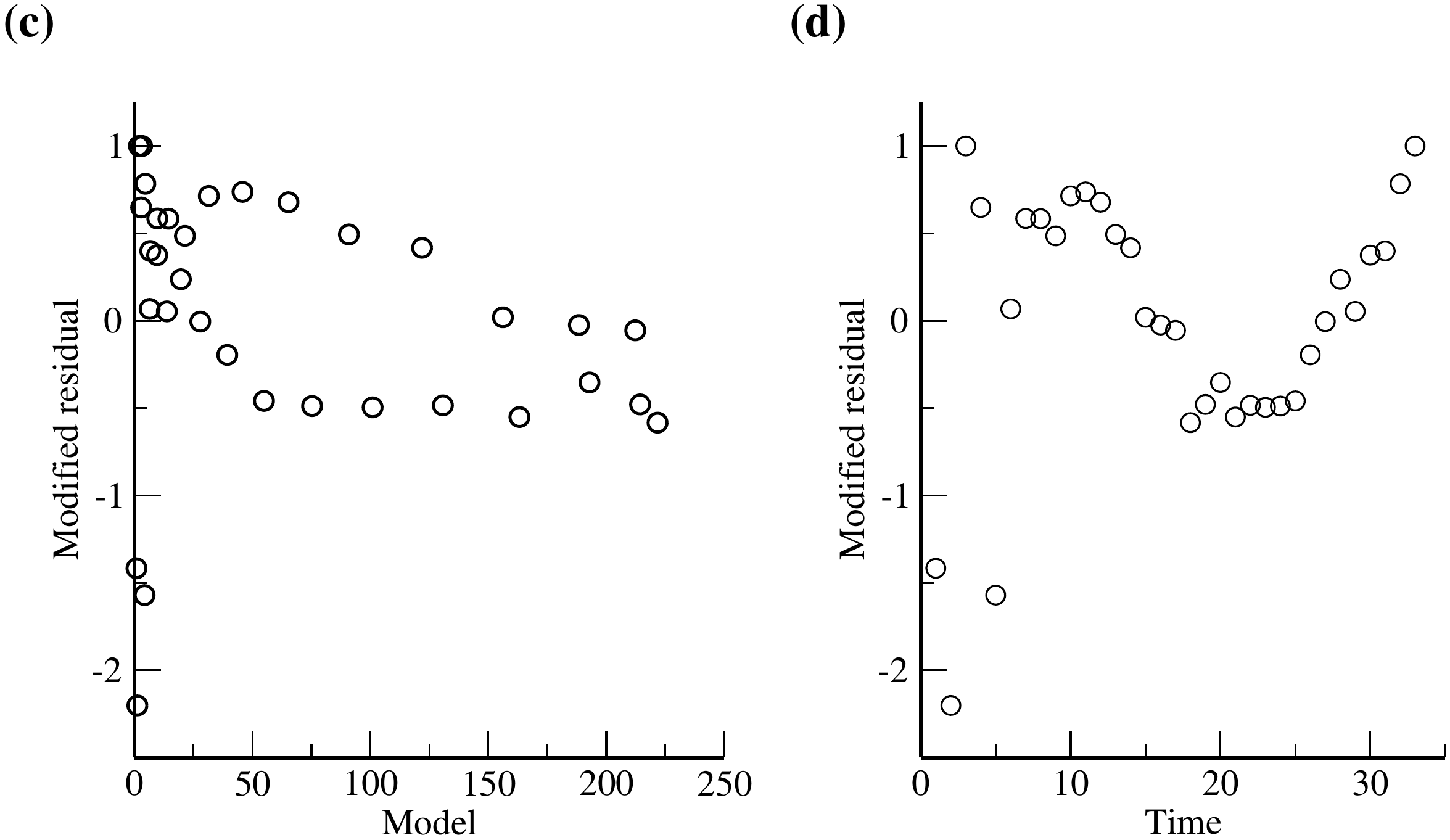}
\caption{GLS applied to influenza data from season 1998-1999. The weights 
were taken equal to $1/z(t_j;\theta)^{2}$.
Panel (a) depicts the observations (solid squares) as well as
the model prediction (solid curve).  In Panel (b) $1,000$ of the $m=10,000$
samples of the effective reproductive number $\mathcal{R}(t)$ are displayed.  
The solid curve depicts the central estimate $\mathcal{R}(t;\hat\theta_{GLS})$
and the dashed curve the median of the $\mathcal{R}(t)$ samples at
each point in time. Panel (c) exhibits the modified residuals 
$(y_j-z(t_j;\hat\theta_{GLS}))/z(t_j;\hat\theta_{GLS})$ plotted versus 
the model predictions, $z(t_j;\hat\theta_{GLS})$. Panel (d) displays the
modified residuals plotted against time.
}
\label{9899_gls}
\end{figure}

\begin{figure}
\centering
\includegraphics[width=5in]{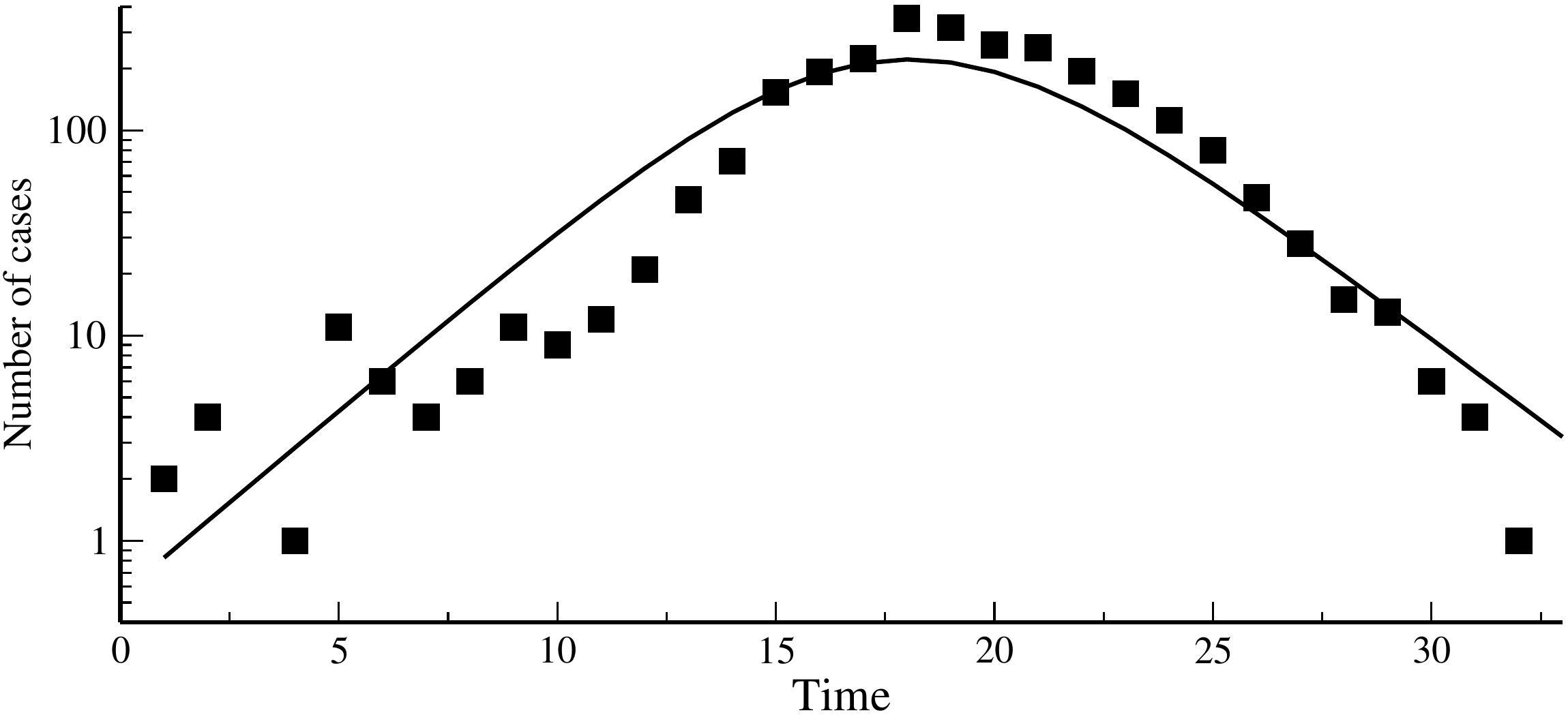}
\caption{Best fitting model for the 1998-1999 season, obtained using GLS 
with $1/z(t_j;\theta)^{2}$ weights. Observations (solid squares) and
the model prediction (solid curve) are plotted on a logarithmic
scale.}
\label{9899_gls_log}
\end{figure}

Visual inspection suggests that the model fits obtained using the GLS approach
 (Figure \ref{9899_gls}) are even worse than those obtained using OLS. 
This is somewhat misleading, however,
because the weights, defined as $w_j=1/[z(t_j;\theta)]^{2}$, mean that
the GLS fitting procedure (unlike visual inspection of the figures) places 
increased emphasis on datapoints whose model value is small and decreased 
emphasis on datapoints where the model value is large. If these graphs
are, instead, plotted with a logarithmic scale on the vertical axis, an
accurate visualization is obtained (Figure \ref{9899_gls_log}): 
multiplicative observation noise on a linear scale becomes 
constant variance additive observation noise on a logarithmic scale.

As before, however, the parameter estimates have standard errors that
are often of the same order of magnitude as the estimates themselves
(Table \ref{9899_gls_table}).
The residuals plots reveal clear patterns and trends (Figure \ref{9899_gls}(c)
 and (d)). Temporal trends in the residuals (and visual inspection of the 
plots depicting the best fitting model and the datapoints) indicate that 
there are systematic differences between the fitted model and the data. 
For instance, it appears that the fitted model peaks slightly earlier than 
the observed outbreak, and, as a result, there are numbers of sequential 
points where the data lies above or below the model. The modified residuals 
versus model plot suggests that the variation of the residuals may be 
decreasing as the model value increases.

The condition number of the matrix
$\chi(\hat\theta_{GLS},n)^T W(\hat\theta_{GLS}) \chi(\hat\theta_{GLS},n)$ is
$9.0\times 10^{19}$. This is very similar to that for the OLS estimation,
again suggesting caution in interpreting the standard errors.

\begin{figure}
\centering
\includegraphics[width=5in]{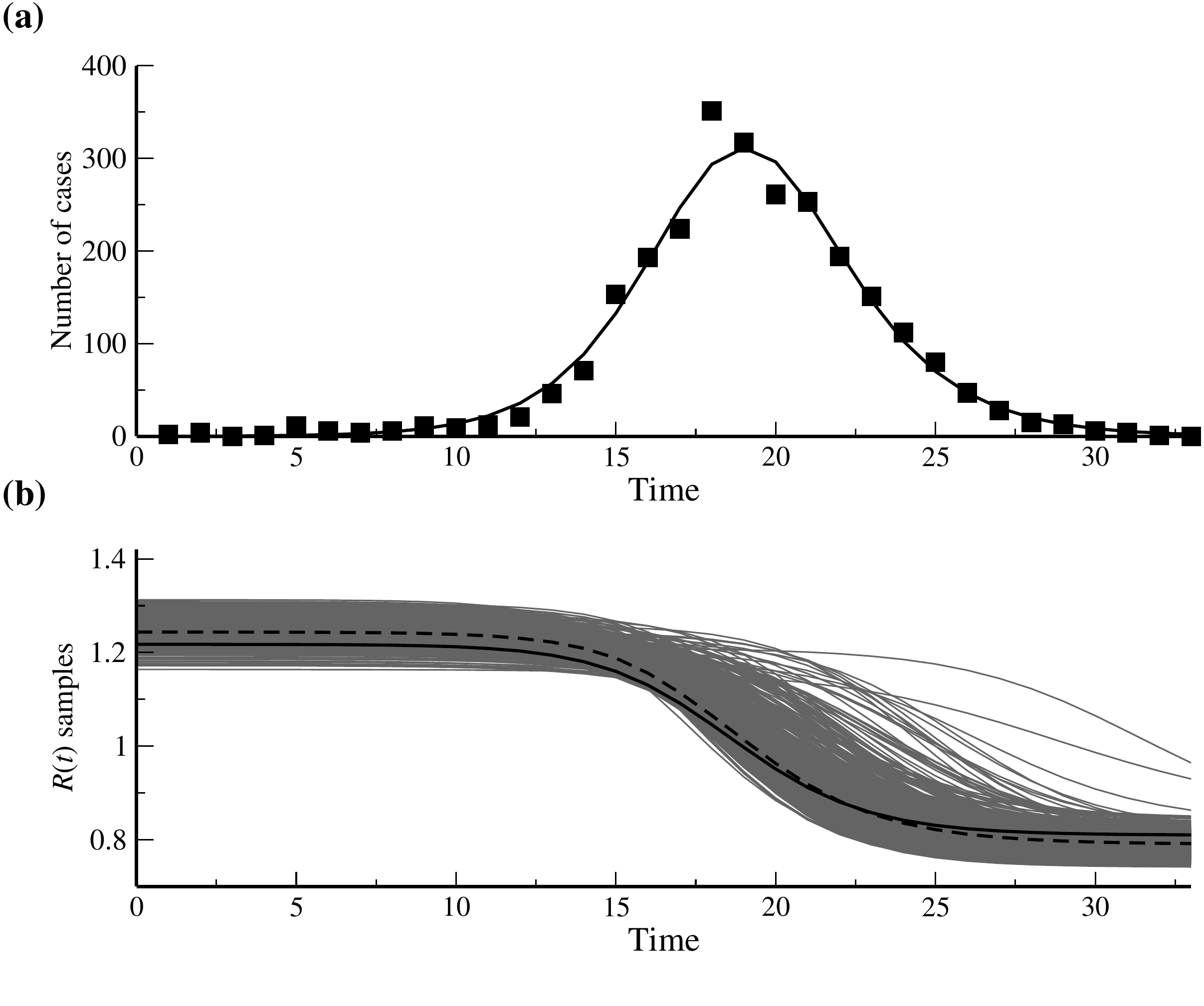}
\includegraphics[width=5in]{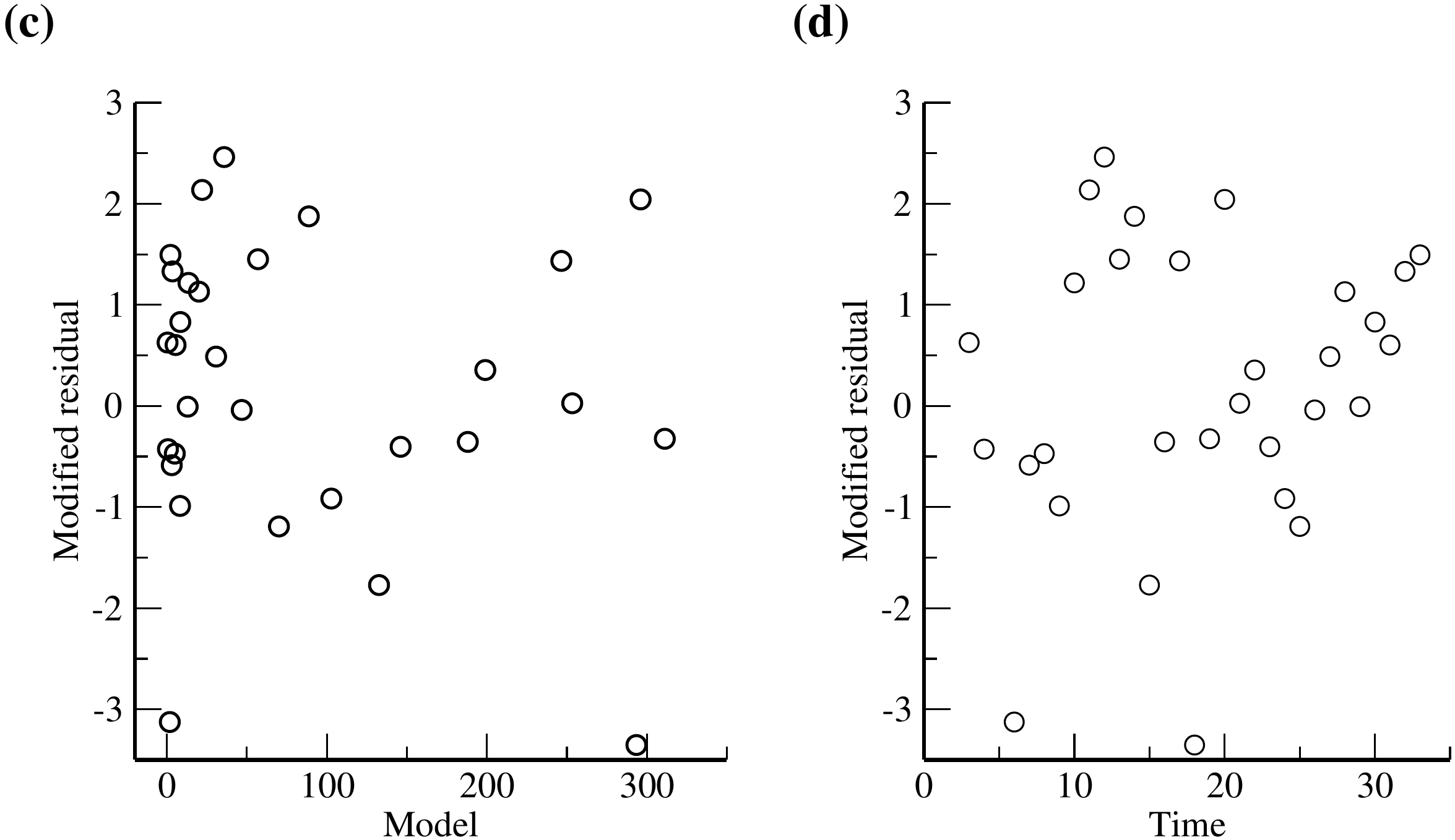}
\caption{GLS estimation for influenza data from season 1998-1999. Weights 
equal $1/z(t_j;\theta)$.
Panel (a) depicts the observations (solid squares) as well as
the model prediction (solid curve).  In Panel (b) $1,000$ of the $m=10,000$
samples of the effective reproductive number $\mathcal{R}(t)$ are displayed,
together with the central estimate $\mathcal{R}(t;\hat\theta_{GLS}$ (solid
curve) and, at each time point, the median of the $\mathcal{R}(t)$ samples
(dashed curve). Panel (c) presents the modified residuals
 $(y_j-z(t_j;\hat\theta_{GLS}))/z(t_j;\hat\theta_{GLS})^{1/2}$ 
versus the model predictions. Panel (d) displays the
modified residuals plotted against time. Three modified residuals fall outside
the range shown on this graph: their values (and the timepoints at which
they arise) are -4.91 (at $t=1$), -7.72 (at $t=2$), and -9.50 (at $t=5$).
}
\label{9899_gls_2}
\end{figure}

\begin{figure}
\centering
\includegraphics[width=6in]{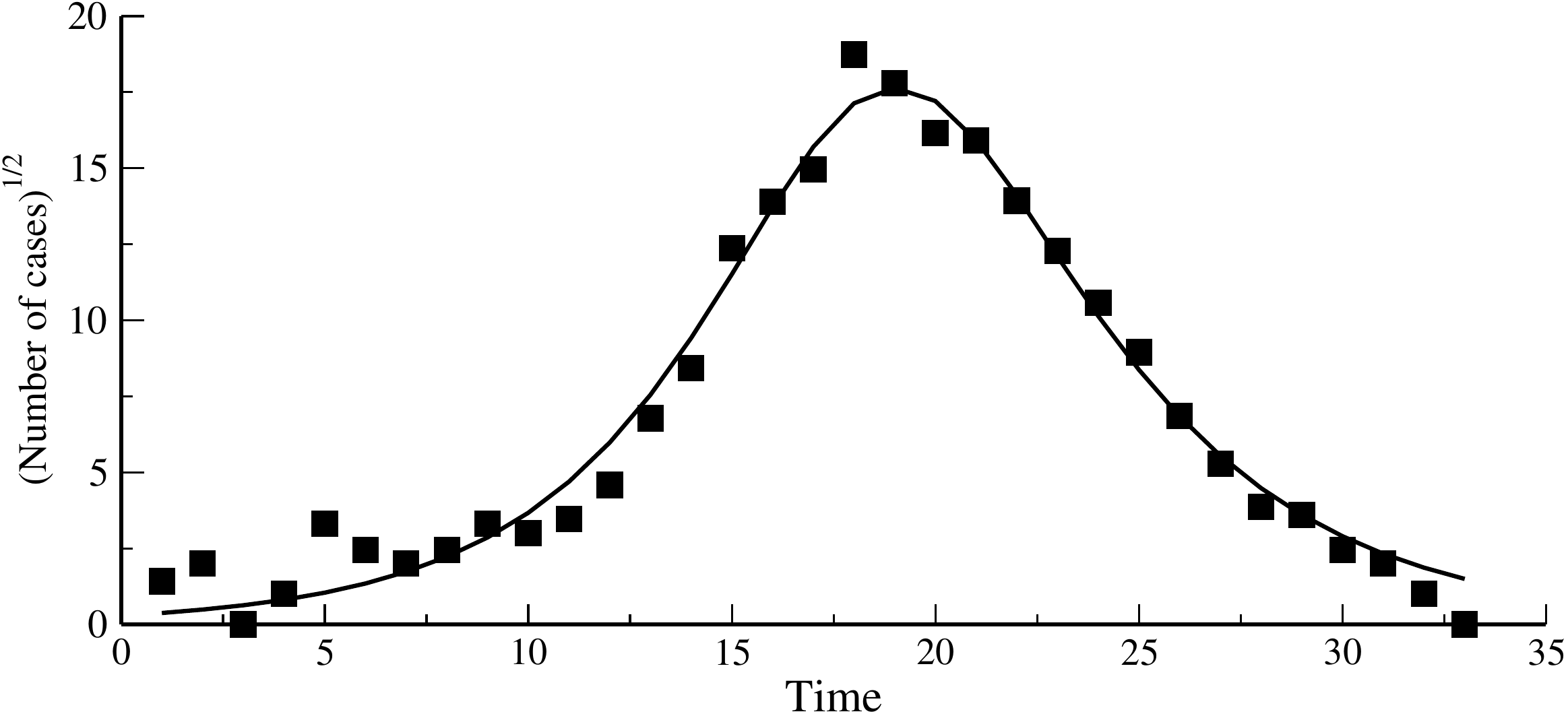}
\caption{Best fitting model for the 1998-1999 season, obtained using GLS 
  with $1/z(t_j;\theta)$ weights. Square roots of observations (solid squares) 
and square roots of the model prediction (solid curve) are plotted.}
\label{9899_gls_sqrt}
\end{figure}

The modified residuals versus model plot indicates that 
the $1/z(t_j;\hat \theta_{GLS})^2$ weights may have overcompensated for the 
non-constant variance seen in the OLS
residuals. This suggests that it may be appropriate to use weights
that vary in a milder fashion, such as $1/z(t_j;\hat \theta_{GLS})$.
The model fits that result with these new weights appear, by
visual inspection, to provide a more satisfactory fit to the data
(Figure \ref{9899_gls_2}). 
Standard errors for the parameter estimates are still large, however 
(Table \ref{9899_gls_2_table}). Our earlier comment concerning the
difficulty of assessing the adequacy of GLS model fits by visual inspection
should be borne in mind--- see Figure \ref{9899_gls_sqrt} for a more
accurate depiction in which square roots of the quantities are plotted
so as to transform the errors in which variance scales with the model value 
into additive errors with constant variance.

The new modified residuals plots, which now focus on the quantities
\[
	\frac{y_j-z(t_j;\hat\theta_{GLS})}{z(t_j;\hat\theta_{GLS})^{1/2}},
\]
also appear to exhibit less marked patterns than they did for either 
OLS or GLS with the $1/z^2$ weights (Figures \ref{9899_gls_2}(c) and (d)).
The condition number for the matrix
$\chi(\hat\theta_{GLS},n)^T W(\hat\theta_{GLS}) \chi(\hat\theta_{GLS},n)$ is 
$1.3\times 10^{20}$, again suggesting ill-posedness and that the standard 
errors should be interpreted with caution.

A handful of surprisingly large modified residuals are seen on 
many occasions, although these often do not appear on our plots because we
choose the range on the residuals axis so that the majority of the
points can be seen most clearly. The locations of these residuals is noted
on the figure caption; we see that they occur during the initial part of
the time series, when the numbers of cases are low.

\begin{table}
\centering
\caption{Results of GLS estimation applied to influenza data from season 1998-1999, weights 
equal to $1/z(t_j;\theta)^{2}$.}
\begin{tabular}{|c|c|c|} \hline\hline
Parameter   & Estimate    &    Standard error\\ \hline
$S_0$&7.939$\times 10^{3}$&1.521$\times 10^{4}$\\ \hline
$I_0$&2.436$\times 10^{-1}$&4.216$\times 10^{-1}$\\ \hline
$\tilde\beta$&3.458$\times 10^{-4}$&5.233$\times 10^{-5}$\\ \hline
$\gamma$&2.333$\times 10^{0}$&5.318$\times 10^{0}$\\ \hline
\multicolumn{3}{|c|}{$L(\hat\theta_{GLS})=$1.754$\times 10^{1}$}\\ 
\multicolumn{3}{|c|}{$\hat\sigma^2_{GLS}=6.047\times 10^{-1}$}\\\hline\hline 
\multicolumn{1}{|c|}{Min. $\mathcal{R}(t;\hat\theta_{GLS})$}&\multicolumn{2}{c|}{0.843\ \ [0.784,1.018]}\\ \hline
\multicolumn{1}{|c|}{Max. $\mathcal{R}(t;\hat\theta_{GLS})$}&\multicolumn{2}{c|}{1.177\ \ [1.052,1.252]}\\ \hline
\end{tabular}
\label{9899_gls_table}
\end{table}

\begin{table}
\centering
\caption{Results of GLS estimation applied to the 1998-1999 season
influenza data. Weights taken to equal $1/z(t_j;\theta)$.}
\begin{tabular}{|c|c|c|} \hline\hline
Parameter   & Estimate    &    Standard error\\ \hline
$S_0$&7.799$\times 10^{3}$&9.269$\times 10^{3}$\\ \hline
$I_0$&3.868$\times 10^{-2}$&3.183$\times 10^{-2}$\\ \hline
$\tilde\beta$&3.643$\times 10^{-4}$&2.760$\times 10^{-5}$\\ \hline
$\gamma$&2.333$\times 10^{0}$&3.462$\times 10^{0}$\\ \hline
\multicolumn{3}{|c|}{$L(\hat\theta_{GLS})=$2.335$\times 10^{2}$}\\ 
\multicolumn{3}{|c|}{$\hat\sigma^2_{GLS}=8.051\times 10^{0}$}\\ \hline\hline
\multicolumn{1}{|c|}{Min. $\mathcal{R}(t;\hat\theta_{GLS})$}&\multicolumn{2}{c|}{0.810\ \  [0.754,0.828]}\\ \hline
\multicolumn{1}{|c|}{Max. $\mathcal{R}(t;\hat\theta_{GLS})$}&\multicolumn{2}{c|}{1.218\ \ [1.200,1.297]}\\ \hline
\end{tabular}
\label{9899_gls_2_table}
\end{table}

\clearpage
\subsection{GLS Estimation Using Truncated Data Sets}\label{glstd}

It is quite plausible that our description of the error structure of the 
data is inadequate when the numbers of cases are at low levels. For instance,
the reporting process might change as the outbreak starts to take hold
(e.g., doctors become more alert to possible flu cases) or comes close
to ending. Also, our model is deterministic whereas a real-world epidemic
contains stochasticity. Stochastic effects may exhibit a 
relatively large impact
at the start or end of an epidemic, when the numbers of cases are low. It
is possible for the infection to undergo extinction, a phenomenon which 
cannot be captured by the deterministic model. Spatial clustering of cases
is also a distinct possibility, particularly during the early stages of
an outbreak. This will affect the time course of an outbreak as well as
the reporting process: clustering of cases may well increase the reporting
noise if cases in a cluster tend to get reported together (e.g., a cluster
occurs within an area where many isolates are sent to the CDC) or not
reported together (e.g., a cluster occurs in an area that has poorer 
coverage in the reporting process). 

Indeed, examination of one of the influenza time series plotted on either a 
logarithmic or square root scale (Figures \ref{9899_gls_log} or 
\ref{9899_gls_sqrt}) indicates that both the start and end of the time series
are problematic. The fit of the model is clearly poorer over these parts
of the time series, which correspond to the times when the observed values
are small.

Both forms of the weights (inversely proportional to the square of the 
predicted incidence or inversely proportional to the predicted incidence)
mean that errors at these small values have considerable impact on the
 cost function, and hence on the GLS estimation process,
although this is less of a concern for the $1/z$ weights.

Another issue
that has been raised by studies of parameter estimation in
biological situations concerns redundancy in information measured when
a system is close to its equilibrium \cite{Banksernst}. 
This might be a relevant issue for the final part of the outbreak data
as there is often a period lasting ten or more weeks when there are few
cases.

\begin{figure}
\centering
\includegraphics[width=5in]{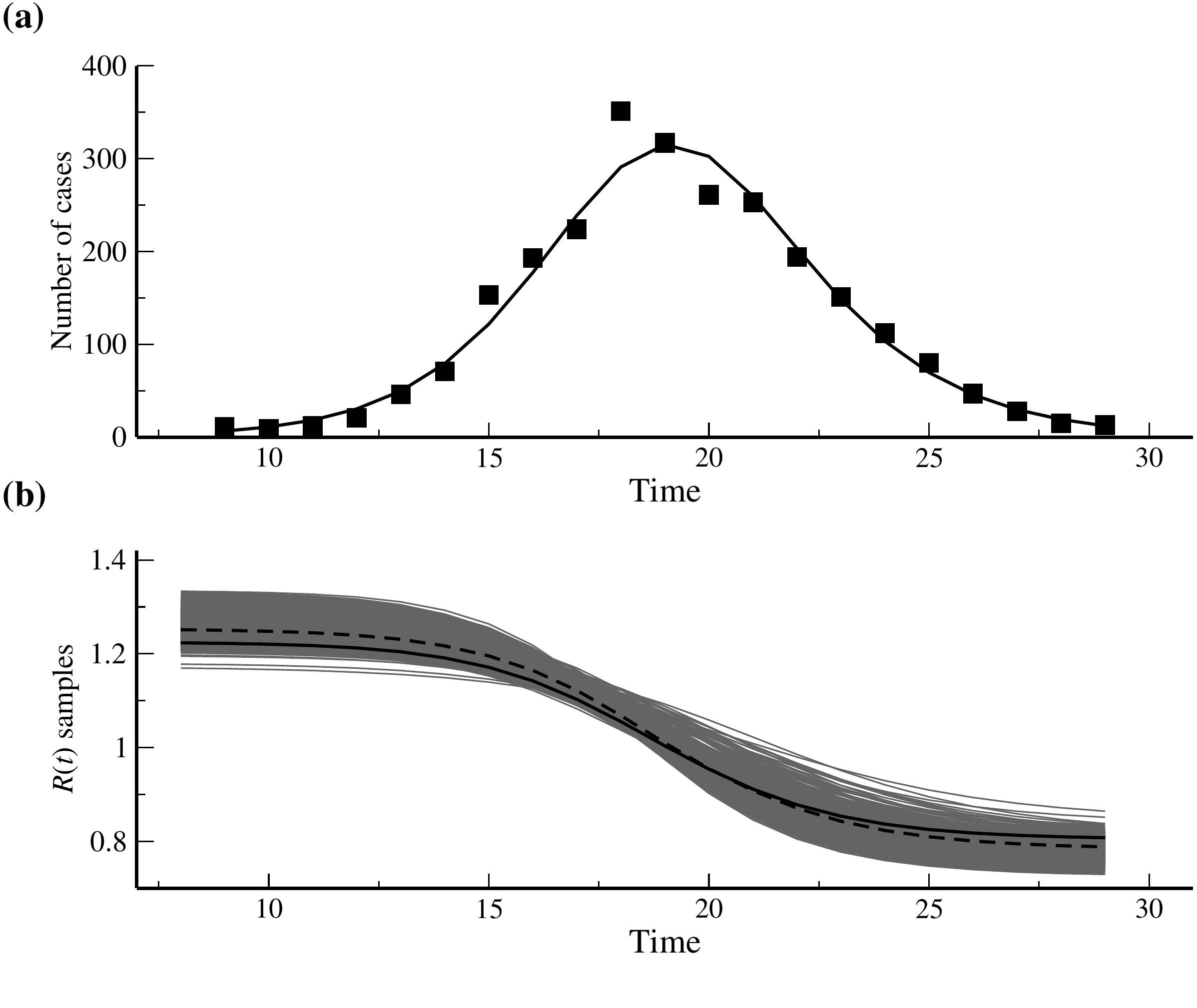}
\includegraphics[width=5in]{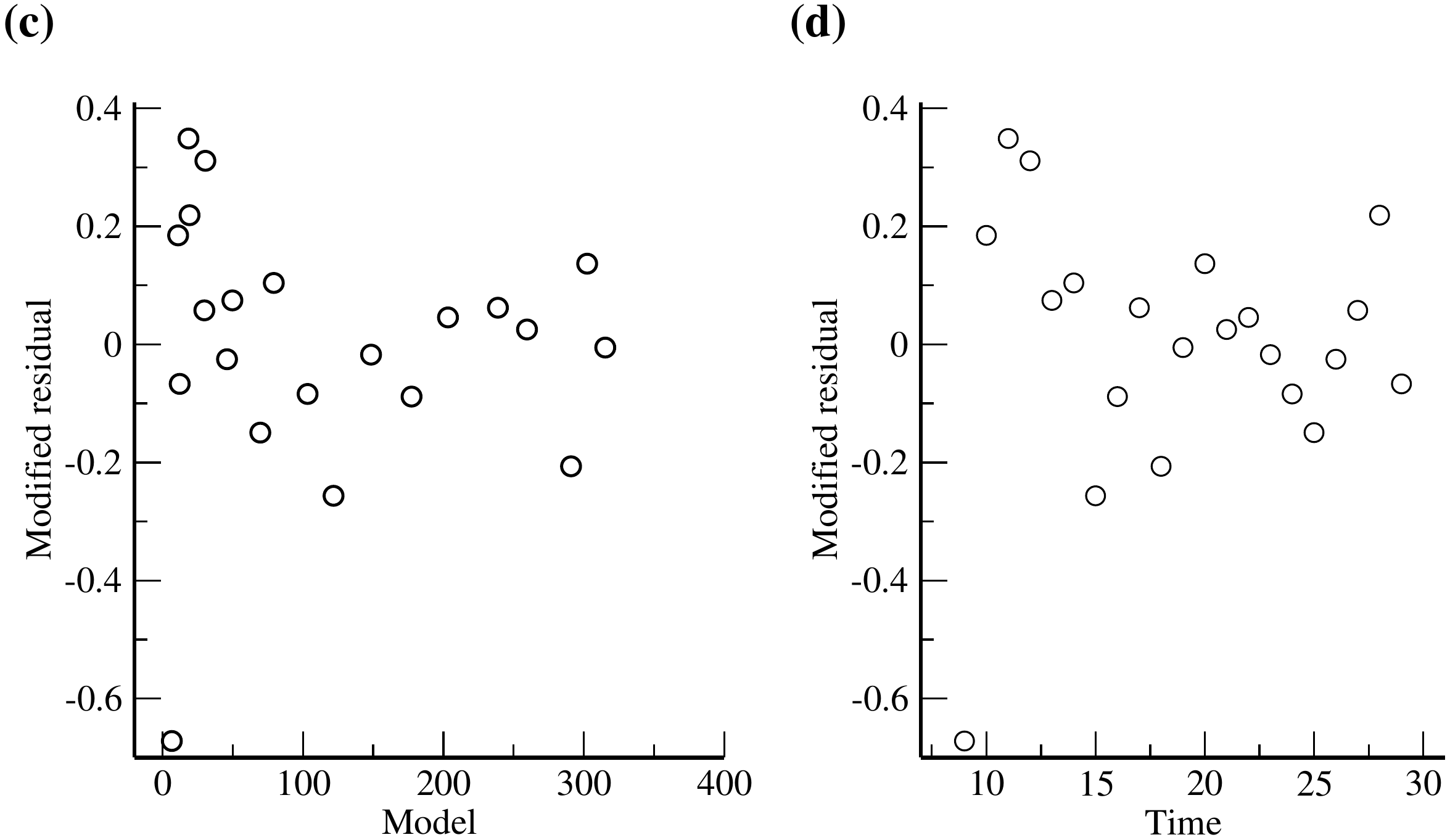}
\caption{Model fits obtained using GLS on truncated influenza data from 
season 1998-1999, weights equal to $1/z(t_j;\theta)^{2}$.
Panel (a) depicts the observations (solid squares) as well as
the model prediction (solid curve).  Panel (b) displays $1,000$ of 
the $m=10,000$ samples of the effective reproductive number, together with
the central estimate $\mathcal{R}(t;\hat\theta_{GLS})$ (solid curve), and
the median of the $\mathcal{R}(t)$ samples at each point in time (dashed
curve). Panel (c) displays the modified residuals versus the model predictions.
In Panel (d) modified residuals are plotted against time.}
\label{9899_gls_trunc}
\end{figure}

\begin{figure}
\centering
\includegraphics[width=5in]{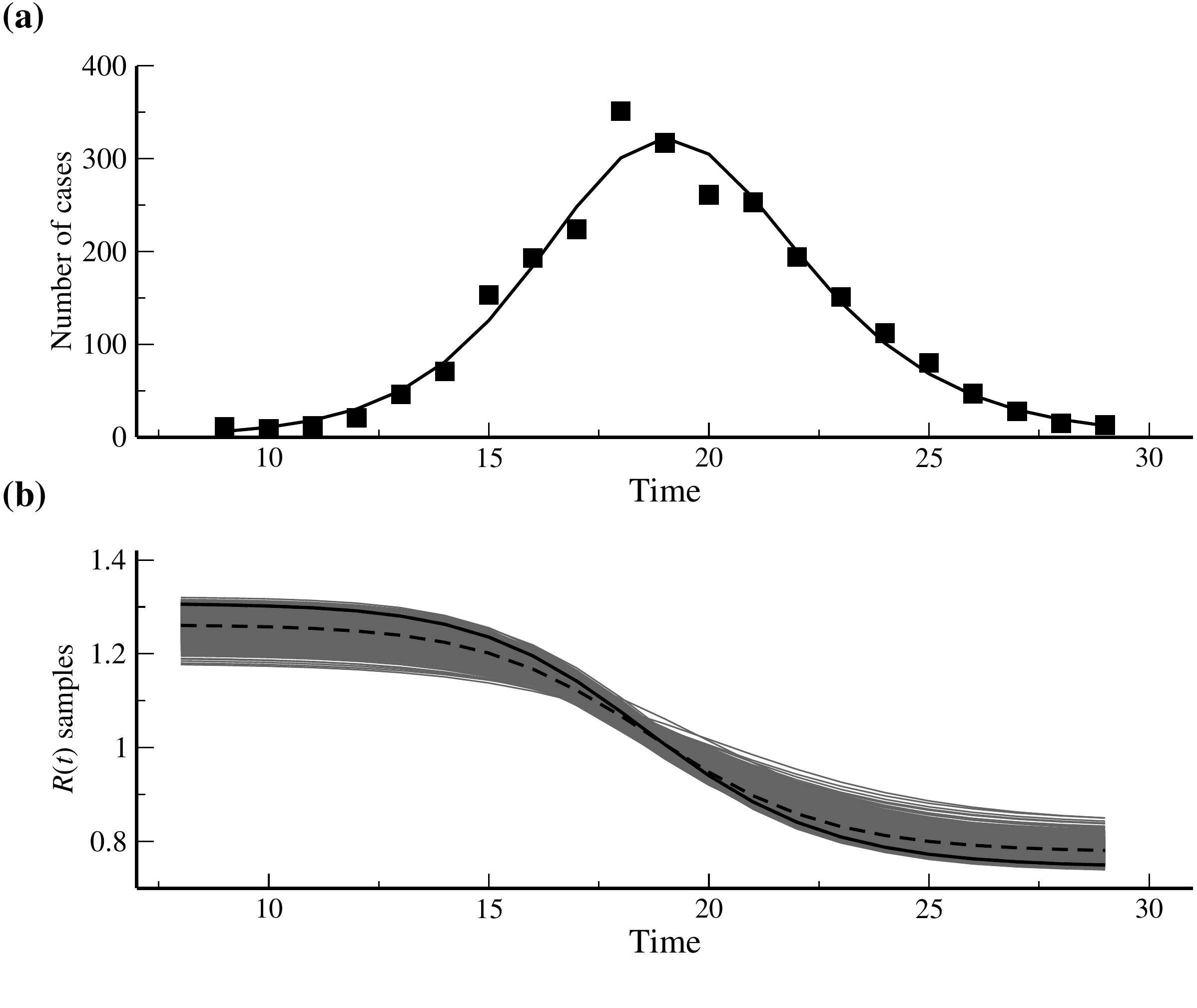}
\includegraphics[width=5in]{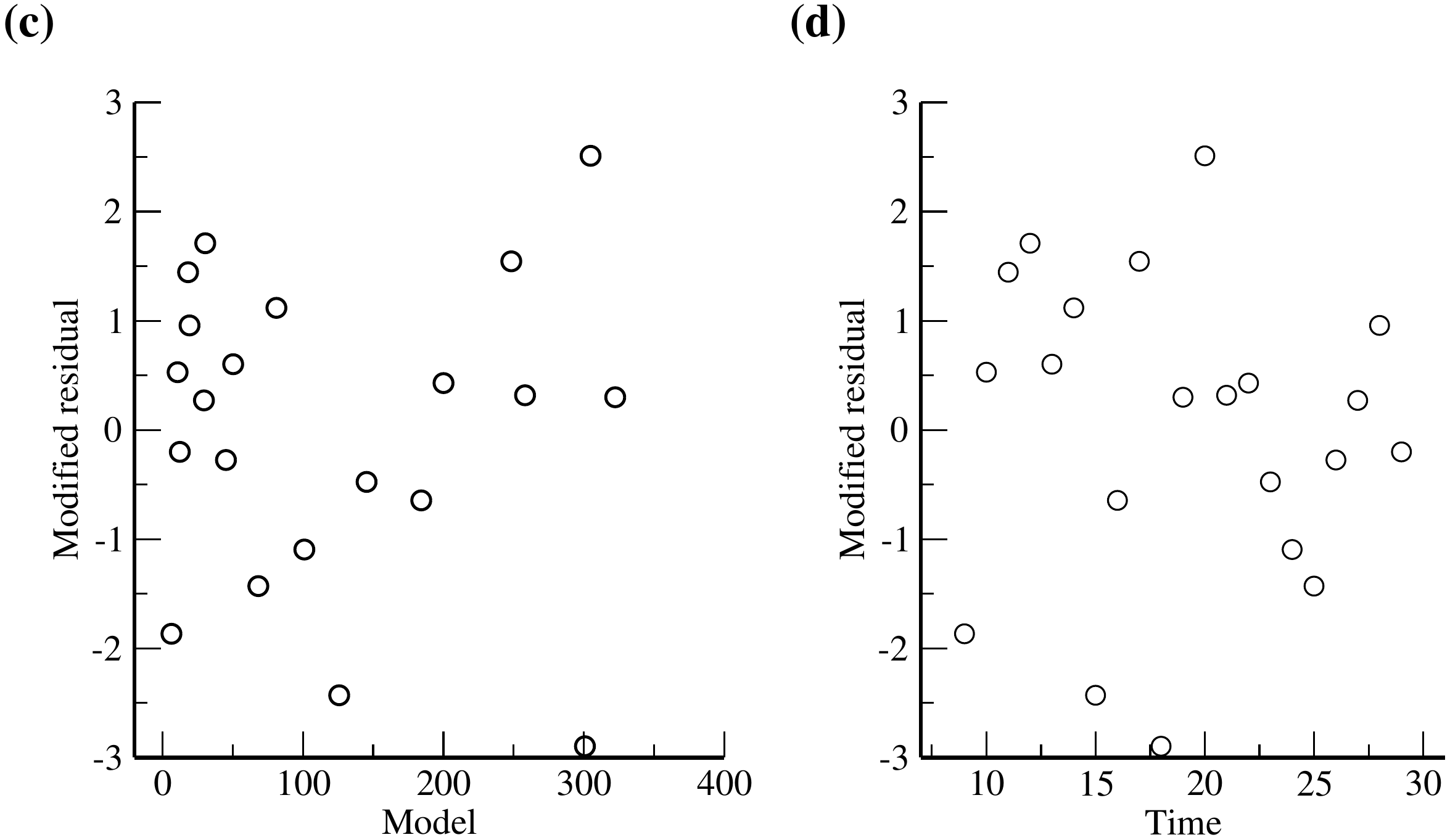}
\caption{
Model fits obtained using GLS on truncated influenza data from 
season 1998-1999, weights equal to $1/z(t_j;\theta)$.
Panel (a) shows the observations (solid squares) as well as
the model prediction (solid curve).  In Panel (b) $1,000$ of the $m=10,000$
samples of the effective reproductive number $\mathcal{R}(t)$ are displayed
together with the central estimate $\mathcal{R}(t;\hat\theta_{GLS})$ 
(solid curve) and the median of the $\mathcal{R}(t)$ samples at each
time point (dashed curve).
Panel (c) shows the modified residuals versus the model prediction.
In panel (d), each modified residual is displayed versus the observation time
point.}
\label{9899_gls_2_trunc}
\end{figure}

We investigated whether the removal of the lowest valued points from the 
data sets would improve the fitting process. We 
constructed truncated data sets by considering only the period between
the time when the number of isolates first reached ten at the start
of the outbreak and first fell below ten at the end of the outbreak.
As a notational convenience, we refer to the numbers of susceptibles
and infectives at the start of the first week of the truncated data set
as $S_0$ and $I_0$, even though these times no longer correspond to
the start of the influenza season. (For example, in Figures 
\ref{9899_gls_trunc} and \ref{9899_gls_2_trunc}, $S_0$ and $I_0$
refer to the state of the system at $t=8$.) 

Comparing Tables \ref{9899_gls_table} and \ref{9899_gls_trunc_table},
which arise from GLS estimation with $1/z^2$ weights, we see that the
standard errors for the parameter estimates have decreased. This decrease
occurs even though the number of points in the data set has fallen from 35 
to 23, causing the factor $1/(n-4)$ that appears in Equation 
(\ref{sigma0_GLS}) to increase by $80\%$. The corresponding
residuals plots (see Figure \ref{9899_gls_trunc}(b) and (c)) 
provide no evidence that the assumptions of the 
statistical model are invalid. The condition number of the matrix
$\chi(\hat\theta_{GLS},n)^T W(\hat\theta_{GLS}) \chi(\hat\theta_{GLS},n)$ 
is $2.4 \times 10^{19}$.

A similar result is seen in the $1/z$ weights case. We remark that we
no longer have the extreme outlier residuals. The condition number of the
matrix $\chi(\hat\theta_{GLS},n)^T W(\hat\theta_{GLS})\chi(\hat\theta_{GLS},n)$
 is $9.2 \times 10^{19}$.

Truncating the data sets has helped considerably with the GLS
estimation process, although the large condition numbers still are cause for 
caution with the standard errors.

Truncation of the data set had little effect on the parameter estimates 
obtained using OLS (results not shown), except that the values of $S_0$ 
and $I_0$ were changed because they refer to a later initial time, as 
discussed above. Standard errors for the OLS estimates were higher than
for the full data set, as should be expected given the reduced 
number of data points.

\begin{table}
\centering
\caption{Estimation results from GLS, with weights  $1/z(t_j;\theta)^{2}$, 
applied to truncated influenza data set for season 1998-1999.}
\begin{center}
\begin{tabular}{|c|c|c|} \hline\hline
Parameter   & Estimate    &    Standard error\\ \hline
$S_0$&7.458$\times 10^{3}$&5.936$\times 10^{3}$\\ \hline
$I_0$&1.758$\times 10^{0}$&1.279$\times 10^{0}$\\ \hline
$\tilde\beta$&3.828$\times 10^{-4}$&2.069$\times 10^{-5}$\\ \hline
$\gamma$&2.333$\times 10^{0}$&2.331$\times 10^{0}$\\ \hline
\multicolumn{3}{|c|}{$L(\hat\theta_{GLS})=$9.475$\times 10^{-1}$}\\ 
\multicolumn{3}{|c|}{$\hat\sigma^2_{GLS}=5.573\times 10^{-2}$}\\ \hline\hline
\multicolumn{1}{|c|}{Min. $\mathcal{R}(t;\hat\theta_{GLS})$}&\multicolumn{2}{c|}{0.808\ \ [0.745,0.820]}\\ \hline
\multicolumn{1}{|c|}{Max. $\mathcal{R}(t;\hat\theta_{GLS})$}&\multicolumn{2}{c|}{1.223\ \ [1.211,1.311]}\\ \hline
\end{tabular}
\end{center}
\label{9899_gls_trunc_table}
\end{table}

\begin{table}
\centering
\caption{Estimation results from GLS, with weights  $1/z(t_j;\theta)$, 
applied to truncated influenza data set for season 1998-1999.}
\begin{center}
\begin{tabular}{|c|c|c|} \hline\hline
Parameter   & Estimate    &    Standard error\\ \hline
$S_0$&6.017$\times 10^{3}$&3.287$\times 10^{3}$\\ \hline
$I_0$&2.091$\times 10^{0}$&9.483$\times 10^{-1}$\\ \hline
$\tilde\beta$&3.797$\times 10^{-4}$&1.774$\times 10^{-5}$\\ \hline
$\gamma$&1.750$\times 10^{0}$&1.317$\times 10^{0}$\\ \hline
\multicolumn{3}{|c|}{$L(\hat\theta_{GLS})=$3.872$\times 10^{1}$}\\ \hline\hline
\multicolumn{3}{|c|}{$\hat\sigma^2_{GLS}=2.277\times 10^{0}$}\\ \hline\hline
\multicolumn{1}{|c|}{Min. $\mathcal{R}(t;\hat\theta_{GLS})$}&\multicolumn{2}{c|}{0.750\ \ [0.748,0.819]}\\ \hline
\multicolumn{1}{|c|}{Max. $\mathcal{R}(t;\hat\theta_{GLS})$}&\multicolumn{2}{c|}{1.306\ \ [1.212,1.308]}\\ \hline
\end{tabular}
\end{center}
\label{9899_gls_2_trunc_table}
\end{table}

\clearpage
\subsection{Estimation for a Reduced Parameter Set}\label{gls3parcd}

The preceding results indicate that there are difficulties in estimating
the parameter $\gamma$, as witnessed by the number of situations in which
the estimate lies on the boundary of the feasible parameter region.
Because $\gamma$ is the one parameter for which we can obtain reasonably
reliable estimates without the need to fit a model to an incidence time 
series \cite{ccvvb,lkmf}, we fix its value and investigate estimation 
for a reduced three parameter problem. In all of what follows, we
apply the estimation methodology to the truncated data sets, as discussed
in the previous section.

We use a fixed infectious period of four days, i.e., 
$1/\gamma=4 {\rm\ days\ }=4/7{\rm \ weeks}$, 
and estimate the parameter vector $\theta=(S_0,I_0,\tilde\beta)$ using
the OLS approach and the GLS approach with weights $w_j=1/[z(t_j;\theta)]^{2}$
or $w_j=1/[z(t_j;\theta)]$. 

Estimation for the reduced parameter set
leads to model fits that are not so different from those obtained using the 
full ($p=4$) set of parameters in $\theta$. For example, for the truncated
data set from the 98-99 season, 
with weights equal to $1/z(t_j;\theta)$, we have $L(\hat \theta_{GLS})=38.72$
for the full parameter set (Table \ref{9899_gls_2_trunc_table}) while  
$L(\hat \theta_{GLS})=38.72$ for the reduced parameter set 
(Table \ref{gls_reduced_9899_2}).
 The standard errors of the parameters, however, are smaller for the reduced 
parameter set: the three standard
errors for the estimates of $S_0$, $I_0$ and $\tilde\beta$ are
$3.287\times 10^3$, $9.483\times 10^{-1}$ and $1.774\times 10^{-5}$,
respectively, for the full parameter set, while they are $2.171\times 10^2$,
$3.174\times 10^{-1}$ and $1.547\times 10^{-5}$ for the reduced set. 

The condition numbers of the matrices
$\chi(\hat\theta_{GLS},n)^T W(\hat\theta_{GLS})\chi(\hat\theta_{GLS},n)$ are
$4.5\times 10^{16}$ and $4.2\times 10^{16}$ for the $1/z^2$ and $1/z$
weights, respectively. 
The increased precision of the estimates here likely results from
identifiability issues in the estimation problem for the full set of model
parameters.

\begin{table}
\centering
\caption{Estimation of three epidemiological parameters. Results obtained
by applying OLS to truncated influenza data set from 
season 1998-1999.} 
\begin{tabular}{|c|c|c|} \hline\hline
Parameter   & Estimate    &    Standard error\\ \hline
$S_0$&6.134$\times 10^{3}$&3.329$\times 10^{2}$\\ \hline
$I_0$&2.442$\times 10^{0}$&5.435$\times 10^{-1}$\\ \hline
$\tilde\beta$&3.707$\times 10^{-4}$&2.348$\times 10^{-5}$\\ \hline
\multicolumn{3}{|c|}{$J(\hat\theta_{OLS})=$6.131$\times 10^{3}$}\\ 
\multicolumn{3}{|c|}{$\hat\sigma^2_{OLS}=3.406\times 10^{2}$}\\ \hline\hline
\multicolumn{1}{|c|}{Min. $\mathcal{R}(t;\hat\theta_{OLS})$ }&\multicolumn{2}{c|}{0.754\ \ [0.744,0.787]}\\ \hline
\multicolumn{1}{|c|}{Max. $\mathcal{R}(t;\hat\theta_{OLS})$ }&\multicolumn{2}{c|}{1.299\ \ [1.258,1.314]}\\ \hline
\end{tabular}
\end{table}
\begin{figure}
\centering
\includegraphics[width=5in]{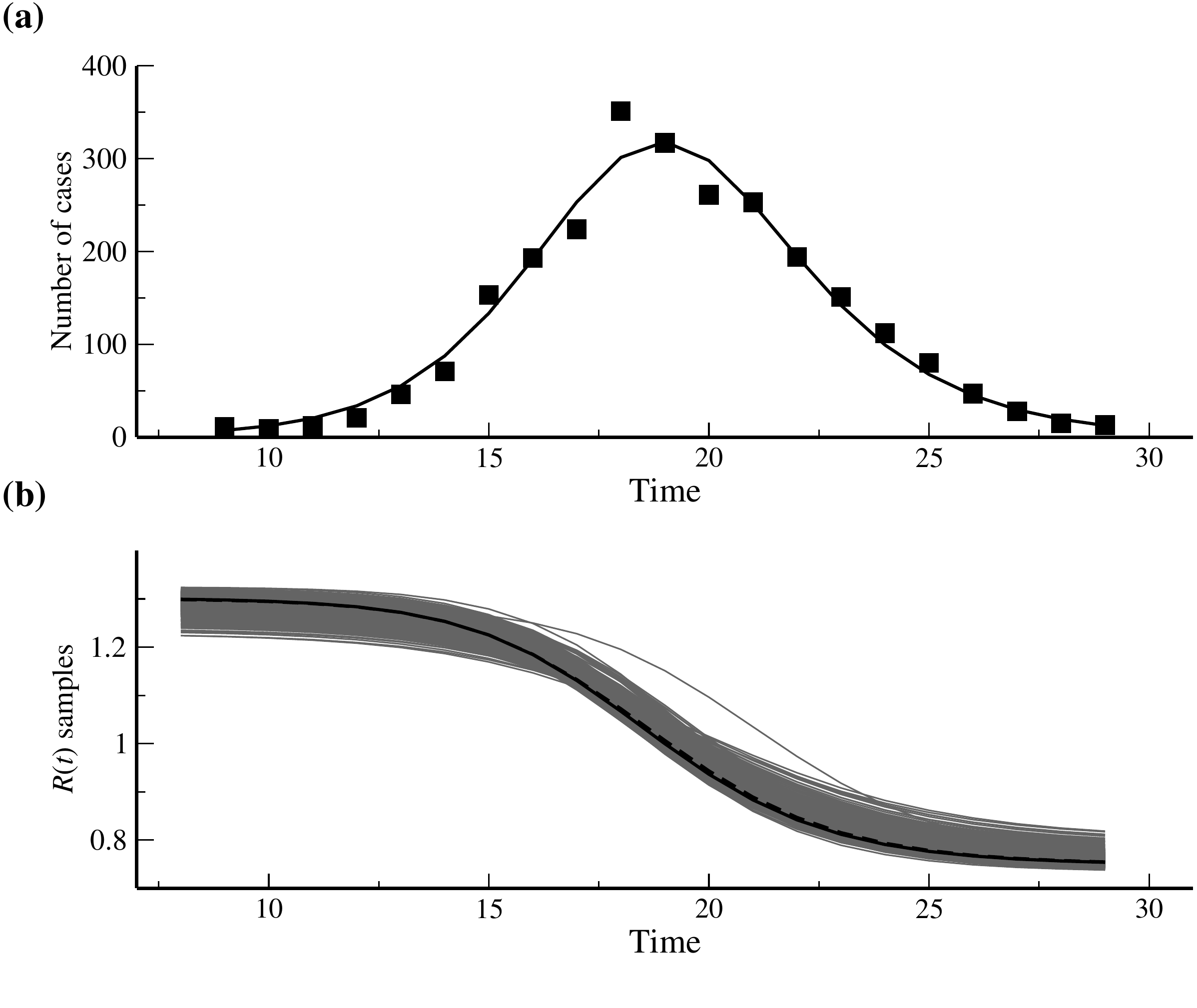}
\includegraphics[width=5in]{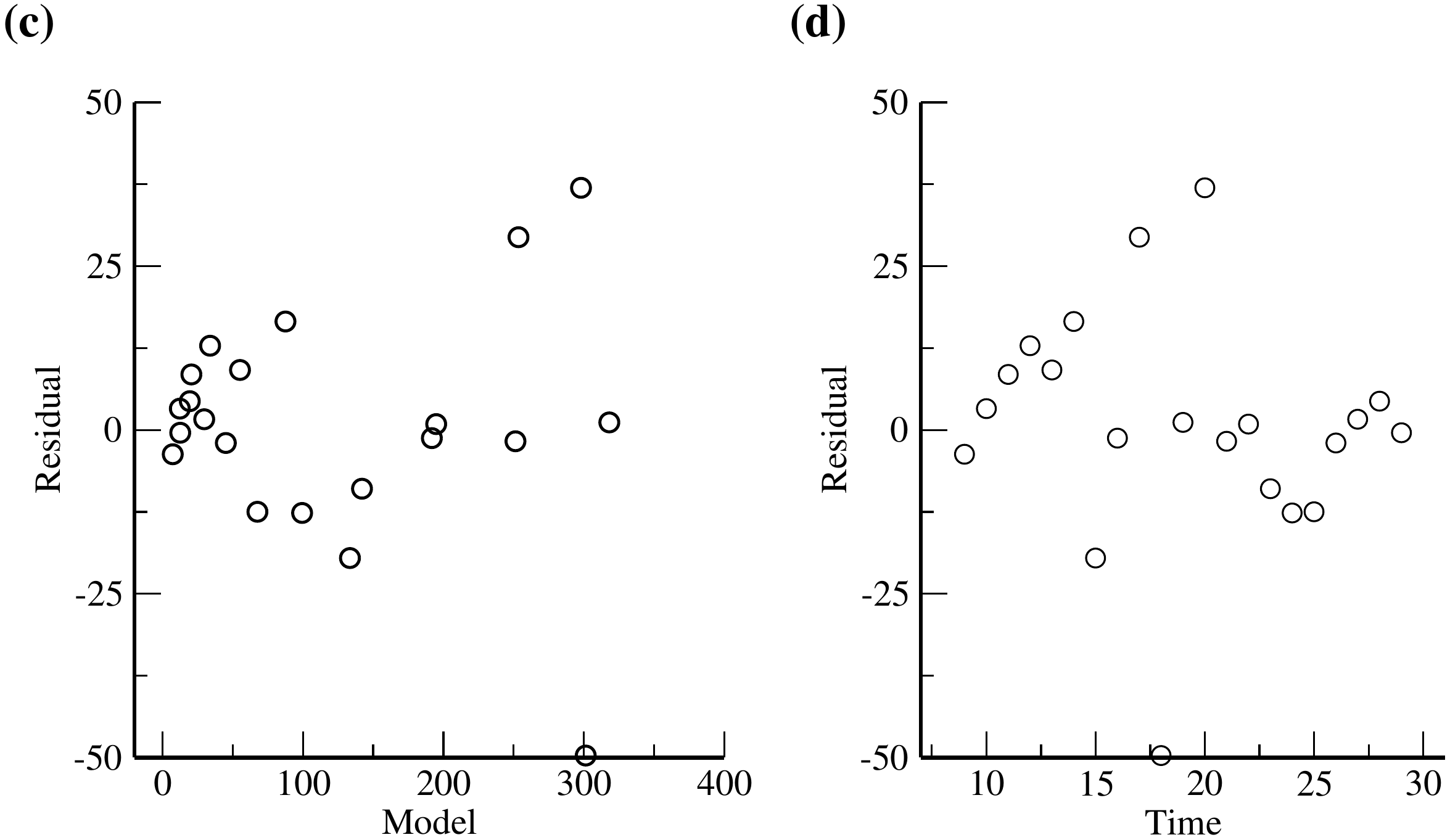}
\caption{Estimation of three epidemiological parameters using OLS on
a truncated data set from the 1998-1999 influenza season.
Panel (a) depicts the observations (solid squares) as well as
the model prediction (solid curve).  In Panel (b) $1,000$ of the $m=10,000$
samples of the effective reproductive number $\mathcal{R}(t)$ are displayed,
together with the central estimate $\mathcal{R}(t;\hat\theta_{OLS})$
(solid curve) and the median of the $\mathcal{R}(t)$ samples at each time
point (dashed curve).  Panel (c) exhibits the residuals 
$y_j-z(t_j;\hat\theta_{OLS})$
versus the model predictions $z(t_j;\hat\theta_{OLS})$. In Panel (d) 
each residual is displayed versus the observation time point 
$t_j=j$, for $j=1\dots,n$.}
\end{figure}

\begin{table}
\centering
\caption{Estimation of three epidemiological parameters. Results obtained
by applying GLS, with 
weights equal to $1/z(t_j;\theta)^{2}$, to truncated influenza data from 
season 1998-1999.} 
\begin{tabular}{|c|c|c|} \hline\hline
Parameter   & Estimate    &    Standard error\\ \hline
$S_0$&$5.985\times 10^{3}$&$3.226\times 10^{2}$\\ \hline
$I_0$&$2.148\times 10^{0}$&2.890$\times 10^{-1}$\\ \hline
$\tilde\beta$&3.808$\times 10^{-4}$&$1.979\times 10^{-5}$\\ \hline
\multicolumn{3}{|c|}{$L(\hat\theta_{GLS})=9.880\times 10^{-1}$}\\ 
\multicolumn{3}{|c|}{$\hat\sigma^2_{GLS}=5.489\times 10^{-2}$}\\ \hline\hline 
\multicolumn{1}{|c|}{Min. $\mathcal{R}(t;\hat\theta_{GLS})$}&\multicolumn{2}{c|}{0.752\ \ [0.740,0.774]}\\ \hline
\multicolumn{1}{|c|}{Max. $\mathcal{R}(t;\hat\theta_{GLS})$}&\multicolumn{2}{c|}{1.302\ \ [1.274,1.320]}\\ \hline
\end{tabular}
\end{table}
\begin{figure}
\centering
\includegraphics[width=5in]{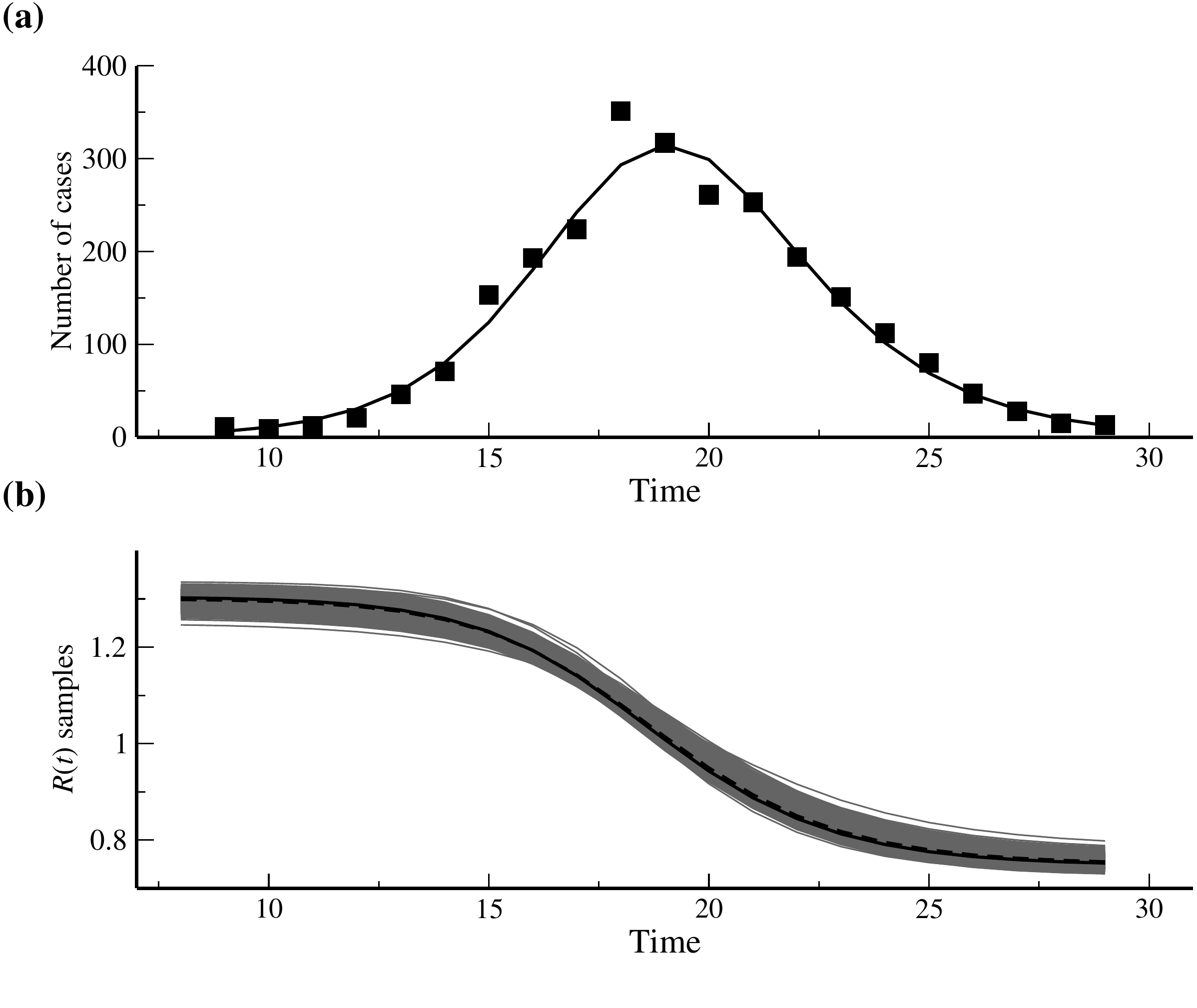}
\includegraphics[width=5in]{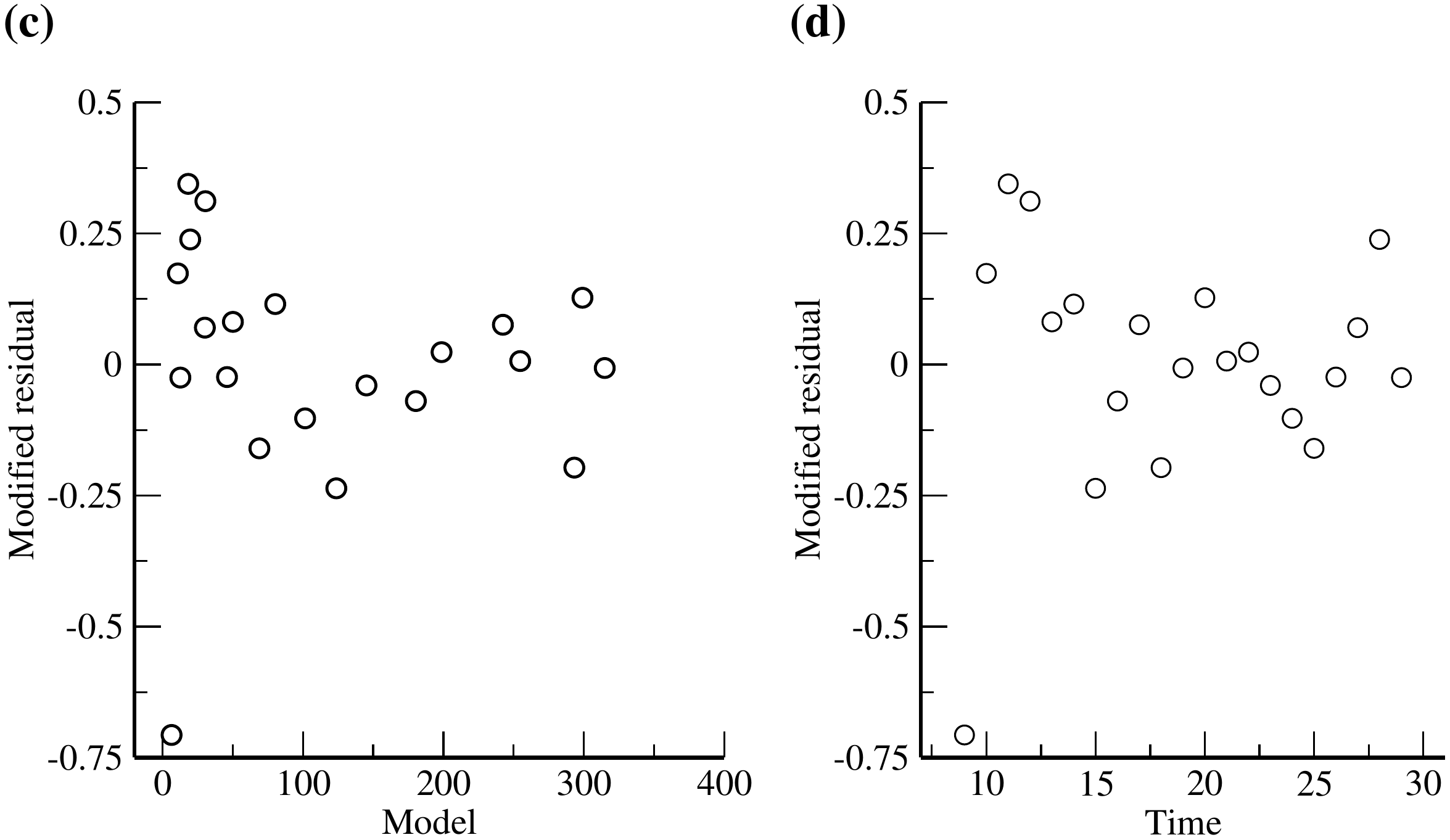}
\caption{Estimation of three epidemiological parameters, using truncated
data from influenza season 1998-1999 and GLS with each weight 
equal to $1/z(t_j;\theta)^{2}$ for $j=1,\dots,n$.
Panel (a) depicts the observations (solid squares) as well as
the model prediction (solid curve).  In Panel (b) $1,000$ of the 
$m=10,000$ samples of the effective reproductive number $\mathcal{R}(t)$ are 
displayed.  The solid curve depicts the central estimate 
$\mathcal{R}(t;\hat\theta_{GLS})$ and, at each time point, the dashed curve 
depicts the median of the $\mathcal{R}(t)$ samples. Panel (c) 
exhibits the modified residuals 
$(y_j-z(t_j;\hat\theta_{GLS}))/z(t_j;\hat\theta_{GLS})$
versus the model predictions $z(t_j;\hat\theta_{GLS})$. In Panel (d) 
each modified residual is displayed versus the observation time point 
$t_j=j$, for $j=1\dots,n$.}
\end{figure}

\begin{table}
\centering
\caption{Estimation of three epidemiological parameters. Results obtained 
by applying GLS, with weights equal to $1/z(t_j;\theta)$, to the truncated
influenza data set from season 1998-1999.}
\begin{tabular}{|c|c|c|} \hline\hline
Parameter   & Estimate    &    Standard error\\ \hline
$S_0$&6.017$\times 10^{3}$&2.171$\times 10^{2}$\\ \hline
$I_0$&$2.090\times 10^{0}$&$3.174\times 10^{-1}$\\ \hline
$\tilde\beta$&3.798$\times 10^{-4}$&1.547$\times 10^{-5}$\\ \hline
\multicolumn{3}{|c|}{$L(\hat\theta_{GLS})=$3.872$\times 10^{1}$}\\ 
\multicolumn{3}{|c|}{$\hat\sigma^2_{GLS}=2.151\times 10^0$}\\ \hline\hline
\multicolumn{1}{|c|}{Min. $\mathcal{R}(t;\hat\theta_{GLS})$}&\multicolumn{2}{c|}{0.750\ \ [0.739,0.767]}\\ \hline
\multicolumn{1}{|c|}{Max. $\mathcal{R}(t;\hat\theta_{GLS})$}&\multicolumn{2}{c|}{1.306\ \ [1.283,1.321]}\\ \hline
\end{tabular}
\label{gls_reduced_9899_2}
\end{table}
\begin{figure}
\centering
\includegraphics[width=5in]{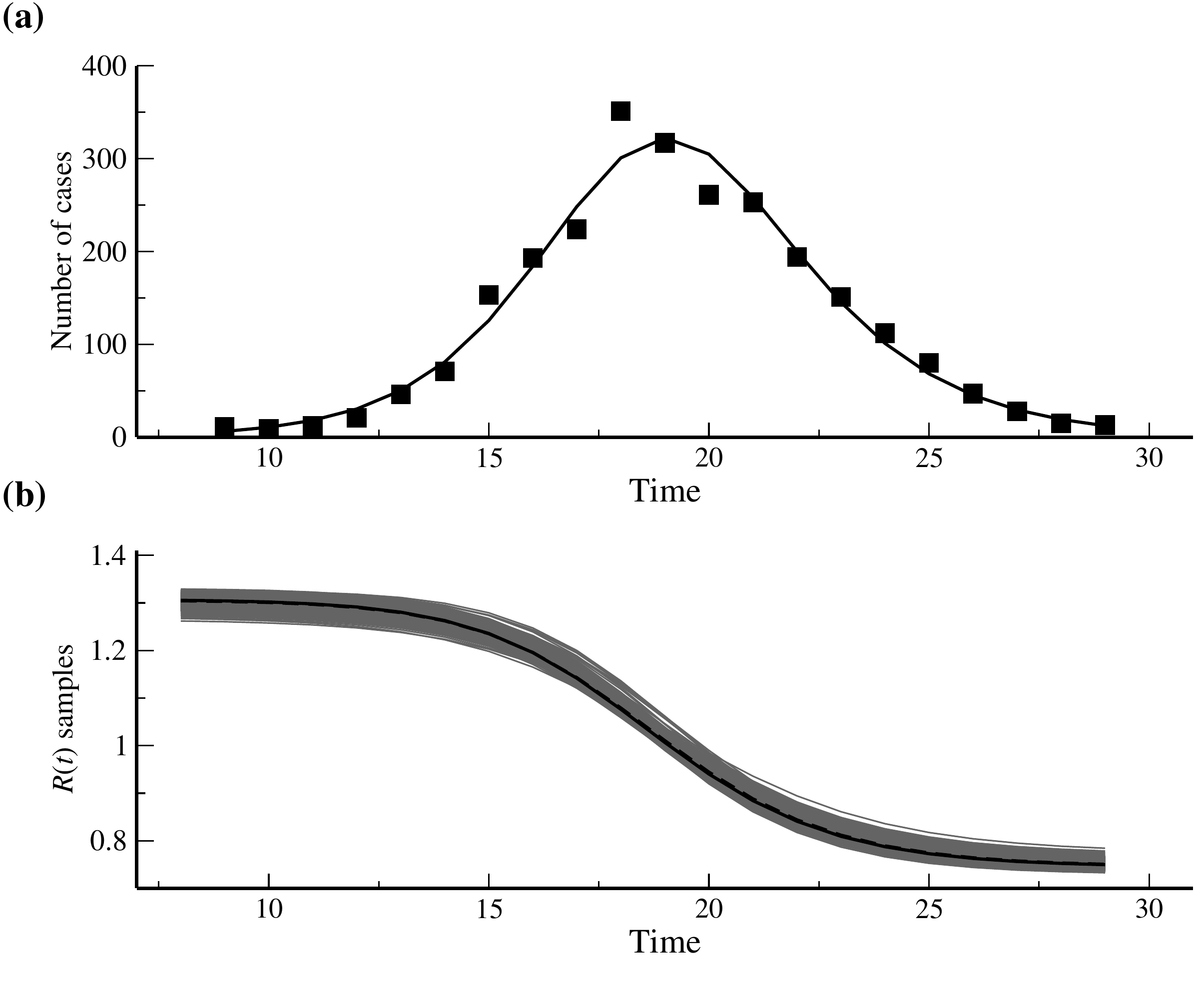}
\includegraphics[width=5in]{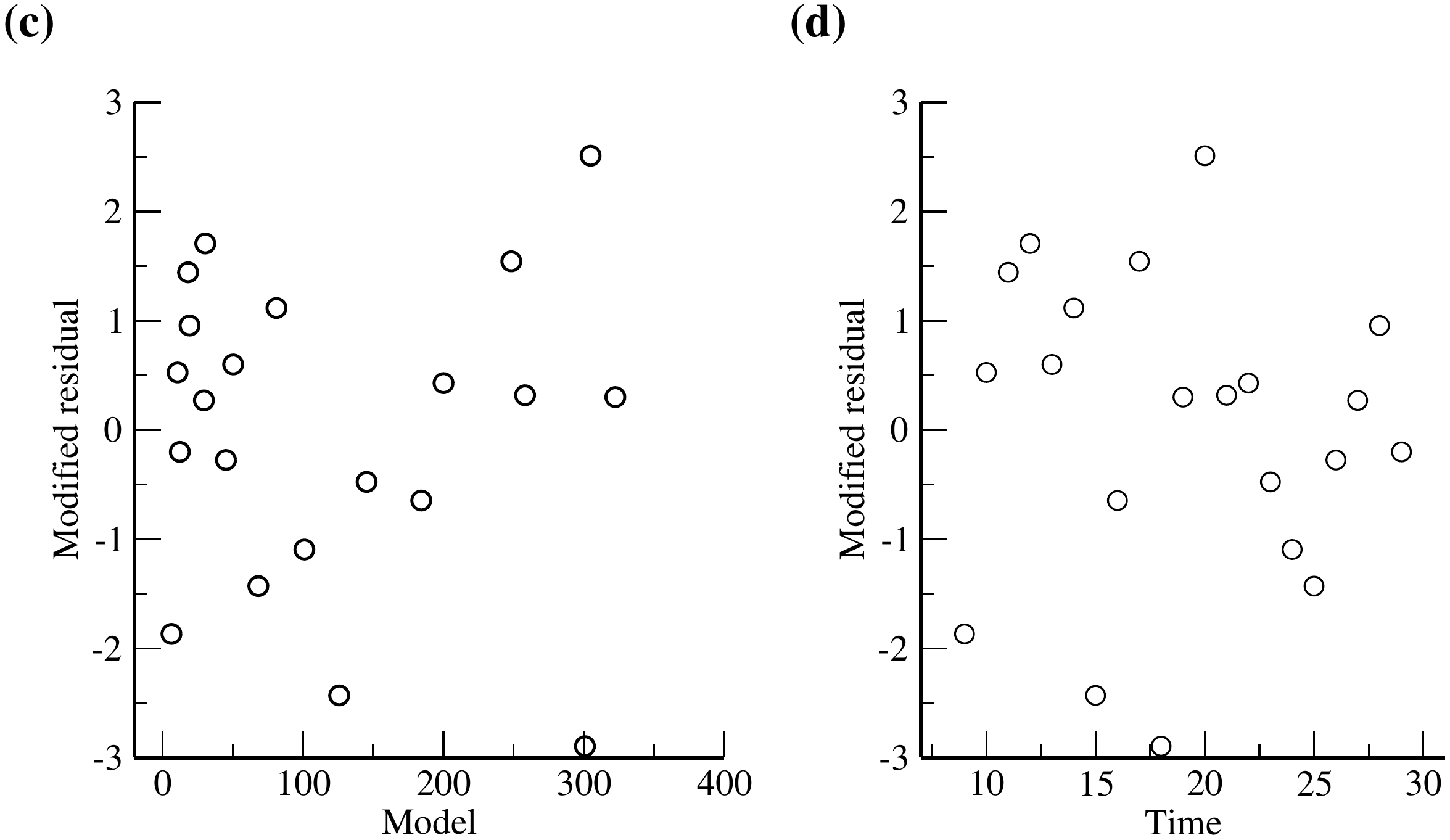}
\caption{Estimation of three epidemiological parameters using truncated
influenza data from season 1998-1999. GLS was used, 
with each weight equal to $1/z(t_j;\theta)$ for $j=1,\dots,n$.
Panel (a) depicts the observations (solid squares) as well as
the model prediction (solid curve).  In Panel (b) $1,000$ of the $m=10,000$
samples of the effective reproductive number $\mathcal{R}(t)$ are displayed.  
Also shown are the central estimate $\mathcal{R}(t;\hat\theta_{GLS})$
(solid curve) together with the median of the $\mathcal{R}(t)$ samples
(dashed curve). Panel (c) exhibits the modified residuals 
$(y_j-z(t_j;\hat\theta_{GLS}))/z(t_j;\hat\theta_{GLS})^{\frac{1}{2}}$
versus the model predictions $z(t_j;\hat\theta_{GLS})$. In Panel (d) 
each modified residual is displayed versus the observation time point 
$t_j=j$, for $j=1\dots,n$.}
\end{figure}

\clearpage
\section{Discussion}\label{diss}

We have presented parameter estimation methodologies that, using 
sensitivity analysis and asymptotic statistical theory, also
provide measures of uncertainty for the estimated parameters. The
techniques were illustrated using synthetic data sets, and it was seen that
they can perform very well with reasonable data sets. Even within the ideal 
situation provided by synthetic data, potential problems of the approach 
were identified. Worringly, these problems were not apparent from inspection 
of the uncertainty estimates (standard errors) alone. However, these problems
were revealed by examination of model fit diagnostic plots, constructed
in terms of the residuals of the fitted model. These results argue strongly 
for the routine use of uncertainty estimation, together with careful
examination of residuals plots when using SIR-type models with surveillance
data.

The statistical methodology presented here only addresses the impact
of observation error on parameter estimation. While the approach can
handle different statistical models for the observation process, it
does assume that we have a model that correctly describes the behavior
of the system, albeit for an unknown value of the parameter vector.
The methodology does not examine the effect of mis-specification of the
model. It is well-known that this effect can dwarf the uncertainty 
that arises from observation error \cite{Lloyd01}. Examination of
residuals plots, however, can identify systematic deviations between
the behavior of the model and the data.

Application of the least squares approaches to the influenza isolate
data gave mixed results. Estimates of the effective reproductive number
were in broad agreement with results obtained in other studies 
(see Table \ref{tabinterpand}). While apparently reasonable fits were 
obtained in some instances, the uncertainty
analyses highlighted situations in which visual inspection suggested
that a good fit had been obtained but for which estimated parameters had
large uncertainties. Residuals plots showed that error variance may not have
been constant (i.e., observation noise was not simply additive), but more
likely scaled according to either the square of the fitted value (i.e., 
relative measurement error) or the fitted value itself.
The potentially large impact of errors at low numbers of cases on
the GLS estimation process was clearly observed.

Temporal trends were observed in the residuals plots, indicative of systematic 
differences between the behavior of the SIR model and the data. Potential
sources of these differences include inadequacies of the model to
describe the process underlying the data and issues with the reliability
of the data itself, particularly in the light of the health warning 
attached to the data by the CDC. (We emphasize, however, that our use of
these data sets should be seen as only an {\it illustration\/} of the 
approach.)

Sophisticated mathematical and statistical algorithms and analyses 
can be utilized to fit SIR-type epidemiological models to surveillance data.
Good quality data is required if this approach is to be successful. In many
instances, however, the available surveillance data is most likely
inadequate to validate the SIR model with any degree of confidence. This
is likely to be true in much of the modeling efforts for epidemics where
the data collection process has inadequacies.

\begin{table}
\caption{Comparison between reproductive number estimates across studies of interpandemic 
influenza.  In this table $\mathcal{R}_0$ stands for the basic reproductive number (naive population),
while $\max(\mathcal{R}(t))$ denotes the initial effective reproductive number in a non-naive population. }
\begin{center}
\begin{tabular}{lr} \hline
Studies of interpandemic influenza & Estimates\\ \hline
Bonabeau et al. \cite{bonatou}& $1.70\leq\mathcal{R}_0\leq3.00$\\
Chowell et al. \cite{chowmillvib}& $1.30\leq \max(\mathcal{R}(t)) \leq1.50$\\
Dushoff et al. \cite{dusplotlevear}& $4.00\leq \mathcal{R}_0\leq16.00$\\
Flahault et al. \cite{flalet} & $\mathcal{R}_0=1.37$\\
Spicer \& Lawrence \cite{spilaw}& $1.46\leq\mathcal{R}_0\leq4.48$\\
Viboud et al. \cite{viboud1}&$1.90\leq \max(\mathcal{R}(t))\leq2.50$ \\ \hline
\end{tabular}		
\end{center}\label{tabinterpand}
\end{table}

\section{Acknowledgments}

This material was based upon work supported by the Statistical and Applied Mathematical
Sciences Institute (SAMSI postdoctoral fellowship to A. C.-A.) which is funded by the National
Science Foundation under Agreement No. DMS-0112069. Any opinions, findings, and conclusions 
or recommendations expressed in this material are those of the authors and do not necessarily reflect the 
views of the National Science Foundation.

\end{document}